\documentclass[journal]{IEEEtran}

\usepackage{duckuments}
\usepackage{graphicx}
\usepackage{subfigure}

\usepackage{chngcntr}   
\usepackage[linesnumbered,ruled,vlined]{algorithm2e}
\newtheorem{Prp1}{\textbf{Proposition}}
\newtheorem{Prf1}{\textbf{Proof}}

\usepackage[final]{changes}

\begin{document}
\title{Efficient Satellite-Ground Interconnection Design for Low-orbit Mega-Constellation Topology}
% a structure at a ~ so this kepdf an author's name from being broken across

\author{Wenhao~Liu,
        Jiazhi~Wu,
        Quanwei~Lin,
        Handong~Luo,
        Qi~Zhang,
        Kun~Qiu,~\IEEEmembership{Senior Member,~IEEE},
        Zhe~Chen,~\IEEEmembership{Member,~IEEE}
        and~Yue~Gao,~\IEEEmembership{Fellow,~IEEE}% <-this % stops a space
\thanks{This work was supported by the National Natural Science Foundation of China under Grant 62341105. \textit{(Corresponding author: Kun Qiu.)}
Wenhao~Liu, Jiazhi~Wu, Quanwei~Lin, Handong~Luo, Qi~Zhang, Kun~Qiu, Zhe~Chen and Yue~Gao are with the School of Computer Science, Fudan University, Shanghai 200438, China, and also with the Institute of Space Internet, Fudan University, Shanghai 200438, China (e-mail: liuwh23@m.fudan.edu.cn; jzwu21@m.fudan.edu.cn; qwlin22@m.fudan.edu.cn; hdluo23@m.fudan.edu.cn; qizhang23@m.fudan.edu.cn; qkun@fudan.edu.cn; zhechen@fudan.edu.cn; gao.yue@fudan.edu.cn).}}% <-this % stops a space

\markboth{IEEE Transactions on Mobile Computing}%
{Wu \MakeLowercase{\textit{et al.}}: Efficient Satellite-Ground Interconnection Design for Low-orbit Mega-constellation Topology}

\maketitle

% modified 10.26 only user terminal
\begin{abstract}
The low-orbit mega-constellation network (LMCN) is an important part of the space-air-ground integrated network system.
An effective satellite-ground interconnection design can result in a stable constellation topology for LMCNs.
A na\"ive solution is accessing the satellite with the longest remaining service time (LRST), which is widely used in previous designs.
The Coordinated Satellite-Ground Interconnecting (CSGI), the state-of-the-art algorithm, coordinates the establishment of ground-satellite links (GSLs). Compared with existing solutions, it reduces latency by 19\% and jitter by 70\% on average.
However, CSGI only supports the scenario where terminals access only one satellite, and cannot fully utilize the multi-access capabilities of terminals. Additionally, CSGI's high computational complexity poses deployment challenges.
To overcome these problems, we propose the Classification-based Longest Remaining Service Time (C-LRST) algorithm.
C-LRST supports the actual scenario with multi-access capabilities. It adds optional paths during routing with low computational complexity, improving end-to-end communications quality.
We conduct our 1000s simulation from Brazil to Lithuania on the open-source platform Hypatia.
Experiment results show that compared with CSGI, C-LRST reduces the latency and increases the throughput by approximately 60\% and 40\%, respectively. 
In addition, C-LRST’s GSL switchings number is 14, whereas CSGI is 23. C-LRST has better link stability than CSGI.
\end{abstract}


% section1 introduction (modified 2023.10.11)

% section1 introduction (modified 2023.10.11)
\section{Introduction}
\label{sec:introduction}
\IEEEPARstart {W}{ith} the continuous expansion of the network scale, low-orbit mega-constellation network (LMCN), has become an important part of the future space-air-ground integrated network due to its wide coverage and all-weather characteristics~\cite{lei2022leo,lin2024fedsn,zhang2024satfed,yuan2024satsense}. LMCN is a satellite network consisting of a series of low-orbit satellites, orbiting at an altitude between 200 and 2,000 km, which can guarantee better signal strength and lower propagation delay~\cite{NTN_in_6G,zhao2024leo,yuan2023graph}. 
The emergence of Starlink~\cite{mcdowell2020low}, Iridium~\cite{curzi2020large}, and other constellations provides examples for the future construction of the space-air-ground integrated network. LMCN provides large-scale Internet access services around the world and has gradually become an important means to achieve globalization, commercialization, and broadband of communications
~\cite{ji_mega_2022,lai_spacertc_2022,luo_refined_2021}. 

As of mid-2021, there are 12,000 satellites of Starlink are already approved by the Federal Communications Commission (FCC), and the total number of Starlink satellites expected to be launched will reach 42,000~\cite{kassas2021enter}.
As the number of satellites increases and the constellation scale expands, the field faces many challenges, including network topology dynamics~\cite{wang2007topological}, radio channel coordination~\cite{al_homssi_next_2022}, routing algorithm design~\cite{DRA}, and ground system design~\cite{guidotti2019lte}. 
% haohao 7.7
% 此外，除了上述关于星座本身的研究，还有一些依赖LMCN的应用的探索。如~\cite{,}研究卫星上机器学习模型的训练效率的优化方法，从而探索利用卫星网络进行高效数据分析的可行性。这些研究专注于卫星网络系统的应用潜力，如对一些观测图像和数据进行分析。
\added{In addition to the above-mentioned research on the constellation itself, there are also some explorations of applications relying on LMCN. For example, some researchers~\cite{SpaceDML,SatelliteFL} study the optimization method for the training efficiency of machine learning models on satellites, thereby exploring the feasibility of using satellite networks for efficient data analysis~\cite{tang2024merit,sun2024efficient}. These studies focus on the application potential of satellite network systems, such as the analysis of observation images and data~\cite{chen2021octopus,lin2024efficient,lin2023pushing,chen2021movi,lin2022channel,tang2024merit,peng2024sums,yang2023efficient}.}

For satellite-ground topology design, link switching and topology updates are frequent owing to the highly dynamic nature of LMCNs.
Taking a satellite at an altitude of 550 km as an example, when the communication angle threshold is 25°, the average switching interval is only 2$\sim$3 minutes~\cite{lei2022leo}. Frequent link switching leads to packet loss and degradation of network performance~\cite{satcp_2023}. An end-to-end throughput measurement for Starlink shows that the throughput standard deviations for Starlink and the terrestrial network are 50.71\% and 34.44\%~\cite{starlink_measurement}, respectively. This instability is likely caused by ground-satellite link (GSL) switchings due to satellite movements.

Algorithms like LRST aim to increase the average service time of satellites and reduce the frequency of GSL switchings, thereby achieving a more stable network topology and enhanced performance. Especially, \added{the} Coordinated Satellite-Ground Interconnecting (CSGI) algorithm performs optimization from a global perspective, minimizing the maximum transmission latency while maintaining stable routing and high network reachability. Compared with traditional methods, CSGI reduces the latency by 19\% and the jitter by 70\% on average~\cite{zhang_enabling_2022}.

Despite the advancements of CSGI algorithms, certain issues remain to be addressed. 
First, CSGI only supports the scenario that terminals can only access a visible satellite, while it cannot fully utilize the multi-access capabilities of the terminals.
However, in actual scenarios, ground stations like those used by Starlink have eight antennas~\cite{FCC}, which means that multiple satellites can be connected to a ground station at the same time. 
% Dish for ordinary users with smaller antennas can also access two satellites, depending on the model of the terminal~\cite{starlink_dish}. 
Second, deploying CSGI in real-world scenarios with hundreds of ground stations presents challenges. Although this algorithm simplifies the handover process, the computational complexity is exponential when initializing the satellite connections to ground stations, which is $O(2^n)$. 

In order to overcome the problems of existing algorithms, such as CSGI, we propose the Classification-based Longest Remaining Service Time (C-LRST) algorithm in this paper.
% C-LRST indudes multi-access into two using a classification strategy, allowing terminals to access two visible satellites simultaneously. 
C-LRST supports the actual scenario of multi-access capabilities of the terminals, and on this basis, adds optional paths during routing with lower computational complexity.
It divides visible satellites into two sets according to their flight directions, and the two access satellites come from two different sets when switching. Based on this classification strategy, the terminal can flexibly select access satellites in different flight directions when routing. 
Also, the core of C-LRST has a lower computational complexity, only $O(n)$, and is easier to deploy to scenarios with hundreds of ground stations than CSGI, whose computational complexity is $O(2^n)$. In our experiments, compared with current algorithms, C-LRST's network delay reduction is about 60\%, the average throughput is also increased by about 40\%, and the link is more stable. Briefly, this paper makes the following contributions:

\begin{itemize}
    \item We study the problem of frequent switching of GSL in the existing satellite-ground topology design and analyse its impacts, such as route loss, frequent rerouting, large network latency fluctuations, and unstable throughput.
    \item By analyzing the distribution of satellites visible to ground terminals, we propose C-LRST, a satellite selection algorithm that is suitable for multi-access capabilities scenarios. In addition, C-LRST is easier to deploy due to its lower computational complexity $O(n)$.
    \item To evaluate the algorithm, we conduct experiments based on \added{Starlink \textit{shell 1}}. 
    The experiment results show that in our $1000s$ long-distance communication simulation from Brazil to Lithuania, compared with CSGI, C-LRST reduces the network latency by about 60\%, and increases the average throughput by about 40\%. In addition, the number of ground-satellite link switchings of C-LRST is 14, whereas CSGI is 23. C-LRST has better link stability than CSGI.
    % 我们通过扩大规模实验来探究C-LRST算法在更大规模星座下的适应性。结果表明，相比基本规模，扩大规模下C-LRST依然保持了较好的性能，这证明了C-LRST算法在密集网络下的可扩展性潜力。此外，我们通过对实验环境和真实场景的对比分析，讨论了C-LRST算法在实际场景中的部署可行性。
    \added{\item We explore the adaptability of C-LRST in a larger-scale constellation by expanding the scale in the experiment. 
    The results show that, compared with the basic scale, C-LRST still maintains good performance under the large scale, which proves the scalability potential of C-LRST in dense networks.
    In addition, through a detailed analysis of the experimental environment and the real world, we discuss the feasibility of C-LRST deployment in actual scenarios.}
\end{itemize}

The rest of the paper is organized as follows. Section~\ref{sec:background} introduces the background and our motivations. Section~\ref{sec:overview} presents the overview design, and our algorithm is explained in Section~\ref{sec:algorithm}. Then, Section~\ref{sec:experiments} describes our simulation experiments and analyses the results. Finally, Section~\ref{sec:conclusion} concludes the whole work.

% section2 background
\section{Background}
\label{sec:background}

\begin{figure}[!htp]
    \centering
    % \begin{subfigure}{0.49\linewidth}
    \subfigure[Kepler's six orbital parameters. The satellite movement can be described by these parameters.]{
        \includegraphics[width=0.46\linewidth]{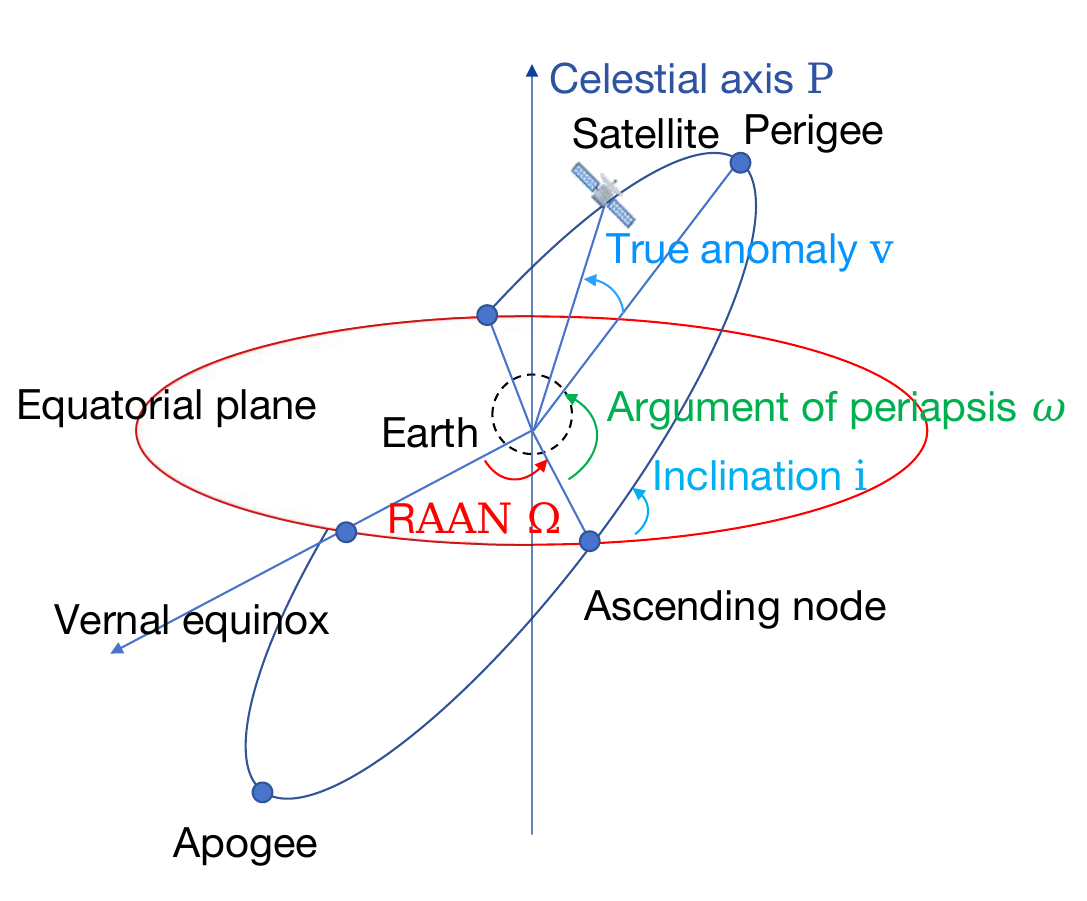}
        %\captionsetup{justification=centering}
        % \caption{Kepler's six orbital numbers}
        \label{fig:Kepler}
    }
    \subfigure[\added{Walker Delta constellation. One satellite establishes four ISLs with adjacent satellites.}]{
        \includegraphics[width=0.46\linewidth]{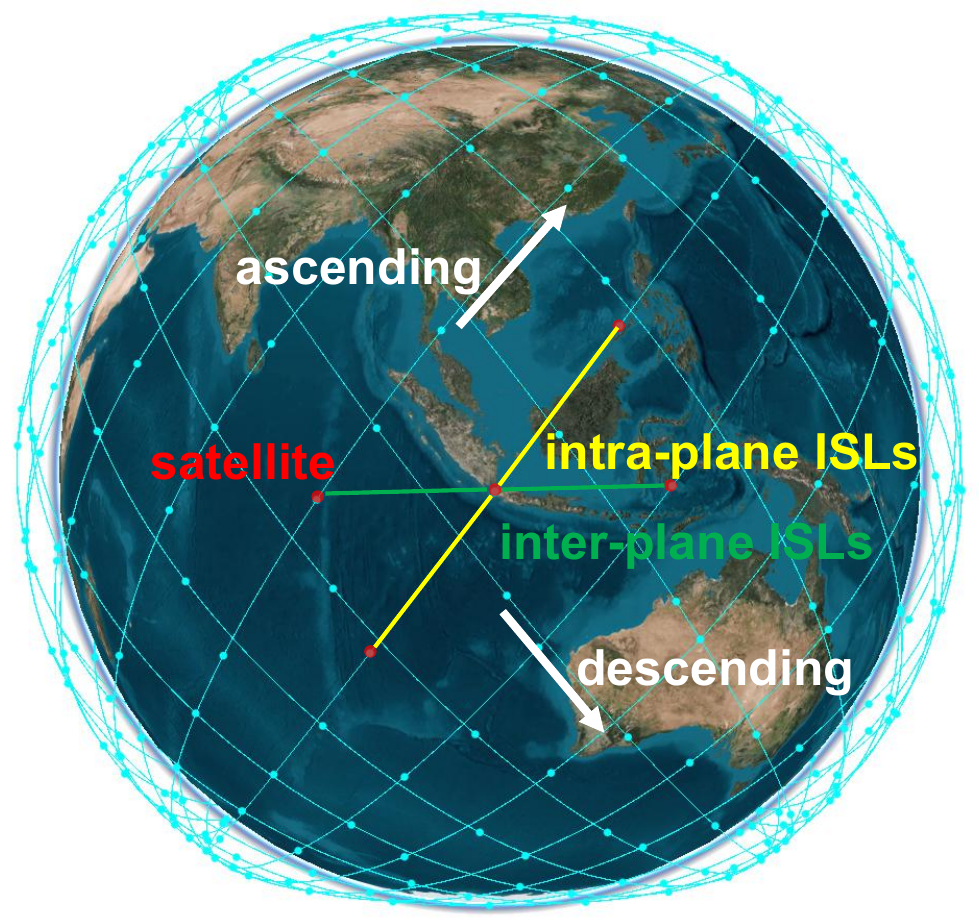}
        \label{fig:walker_delta}
    }
    \subfigure[\added{Elevation angle of satellite. When a satellite's elevation angle exceeds the minimum limit, it is visible relative to the ground station.}]{
        \includegraphics[width=0.6\linewidth]{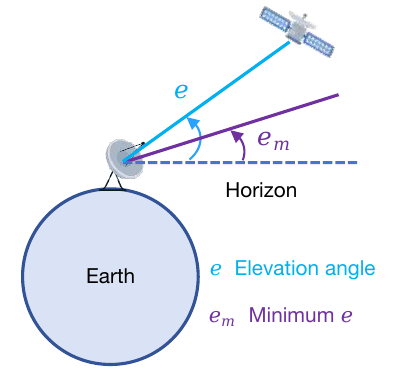}
        \label{fig:elevation}
    }
    \caption{Satellite and Constellation.}
    \label{fig:satellite_and_constellation}
\end{figure}

\begin{figure*}[htb]
    \centering
    \includegraphics[width=0.7\linewidth]{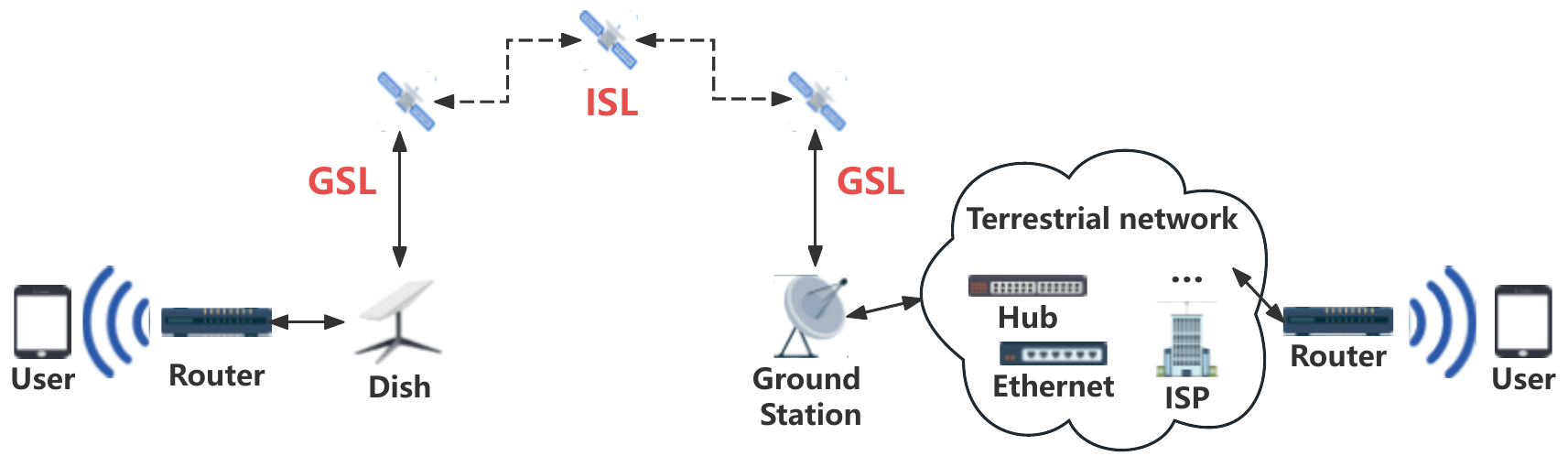}
    %\captionsetup{justification=centering} % 设置图标题居中
    \caption{ISL mode of communication between ground terminals. One ground terminal (such as a user Dish) communicates with the other (such as a ground station) through only two GSLs, with multi-hop ISLs in between.}
    \label{fig:dish_to_gs}
    \hfill
\end{figure*}

\subsection{Constellation, Links and Communications}
\label{sec:background-info}
The satellite movement can be described by Kepler's six orbital parameters, as shown in Fig.~\ref{fig:Kepler}. Internationally, the two-line orbital element (TLE)~\cite{TLE_data} is generally used to describe the satellite ephemeris, which can be easily generated according to the Kepler parameters. 

In this study, we focus on the Walker Delta constellation~\cite{walker_delta}, as shown in Fig.~\ref{fig:walker_delta}. It can be described by four parameters: $N/M/F/\alpha$. $N$ is the number of \added{orbits} in the constellation, $M$ is the number of satellites in the same \added{orbit}, $\alpha$ is the orbit inclination, and $F$ is the phase factor used to adjust the phase difference $\Delta f $ between two satellites in adjacent orbits with the same number in the orbit. $\Delta f = 2\pi F / (N*M)$ and $F$ belongs to the following range: $\{1-N, 2-N, \dots, 0, 1\dots, N-1 \}$. A constellation usually has multiple layers of orbits. This study only considers the layer of Starlink with an altitude of 550km, called \textit{shell 1}.

The inter-satellite link (ISL) adopts the +Grid structure~\cite{+grid}. Under this structure, a satellite can establish four ISLs, two connecting adjacent satellites in the same orbit, and two connecting satellites with the same number in adjacent orbits. \added{The ground-satellite link (GSL) describes the communication between the ground station and the satellite.}

In current satellite networks, there are three main methods of communication between ground terminals \added{(such as ground stations and user Dishes)}.
\begin{itemize}
    \item \added{Transparent forwarding mode: There are no ISLs between satellites, and the communication between ground terminals is relayed through multiple GSLs.}
    \item ISL mode: As shown in Fig.~\ref{fig:dish_to_gs}, \added{one ground terminal communicates with the other through only two GSLs, with multi-hop ISLs in between.}
    \item Hybrid mode: Ground terminals can communicate through multiple \added{ISLs and GSLs}.
\end{itemize}

Transparent forwarding mode is widely used in medium and high orbit constellations~\cite{transparent_forwarding}. For LMCNs, the disadvantage of using Transparent forwarding mode is that it is difficult to achieve communication across continents and oceans and frequent switching of GSLs. 

Starlink plans to deploy large-scale laser ISLs in the future~\cite{starlink_laserlink}, and our \added{satellite-ground interconnection scheme} will adopt the ISL mode, as shown in Fig.~\ref{fig:dish_to_gs}. The Hybrid mode has the shortest routing path because ground nodes are added to the network topology, but similar to the transparent forwarding mode, the network performance is greatly affected by GSL switching. We will analyse this effect through experiments in Section~\ref{sec:experiments} to illustrate the advantages of ISL mode over Hybrid mode.

\subsection{Access satellite and GSL switchings}
As shown in Fig.~\ref{fig:elevation}, $e$ denotes the elevation angle of the satellite relative to the ground station. When its size exceeds the minimum elevation angle $e_m$ of the ground station set by the system, the satellite is visible relative to the ground station~\cite{cakaj_elevation_2014}. When a visible satellite establishes a GSL with a ground station, it is called the access satellite of the ground station. 
% A ground station can have one or more access satellites, depending on its design.

The number of satellites a terminal can access depends on its design. Ground stations like those used by Starlink have eight antennas~\cite{FCC}, which means that multiple satellites can be connected to a ground station at the same time. 
% Dish for ordinary users with smaller antennas can also access two satellites, depending on the model of the terminal.

The altitude range of low Earth orbit (LEO) satellites is 500-1500 kilometers, and the moving speed is approximately 8 km/s~\cite{uzunalioglu_footprint_1999}. For ground terminals, the satellite is visible for around 2-3 minutes~\cite{SpaceRTC}. This means that GSL switchings are happening all the time, leading to network instability and frequent route changes.

% 11.23
\subsection{Classic solutions and issues}
Na\"ive satellite-ground interconnection methods can be summarized into three categories:
\begin{itemize}
\item Longest remaining service time priority when switching (LRST): \added{The ground stations} connect to the satellite with the longest estimated remaining service time until it moves out of range.
\item Nearest distance priority when switching (ND): \added{The ground stations} connect to the nearest satellite until it moves out of range.
\item Always nearest distance priority (AND): \added{The ground stations} always connect to the nearest satellite (i.e., immediately switch to a nearer satellite when it becomes available).
\end{itemize}

According to Zhang et al.~\cite{zhang_enabling_2022}, in LMCN, when the ground station can access only one satellite at the same time, the performance of LRST in terms of RTT, network throughput, etc., is much better than that of ND and AND. Therefore, this paper uses the LRST algorithm as one of the baselines for comparison. 

In addition, we also use the state-of-the-art algorithm, CSGI, for comparison. This algorithm is suitable for situations where the ground station can access only one satellite. When the ground station selects the access satellite, it not only considers the remaining service time but also minimizes the overall latency, thereby achieving global network optimization.

However, although the algorithm has low complexity during the handover, the computational complexity is exponential when initializing the satellite connected to the ground station, which is $O(2^n)$. Therefore, in a scenario with hundreds of ground stations, it cannot be deployed normally for comparison with the other algorithms. Considering this shortcoming, according to the standards of Zhang's paper, we select 15 ground stations with long distances from each other as the ground station set of CSGI algorithm. In the subsequent end-to-end path analysis experiment, the ground station pairs we use are selected from these 15 ground stations to ensure the effectiveness of the comparison experiment.

\begin{figure*}[htb]
    \centering
    \includegraphics[width=\linewidth]{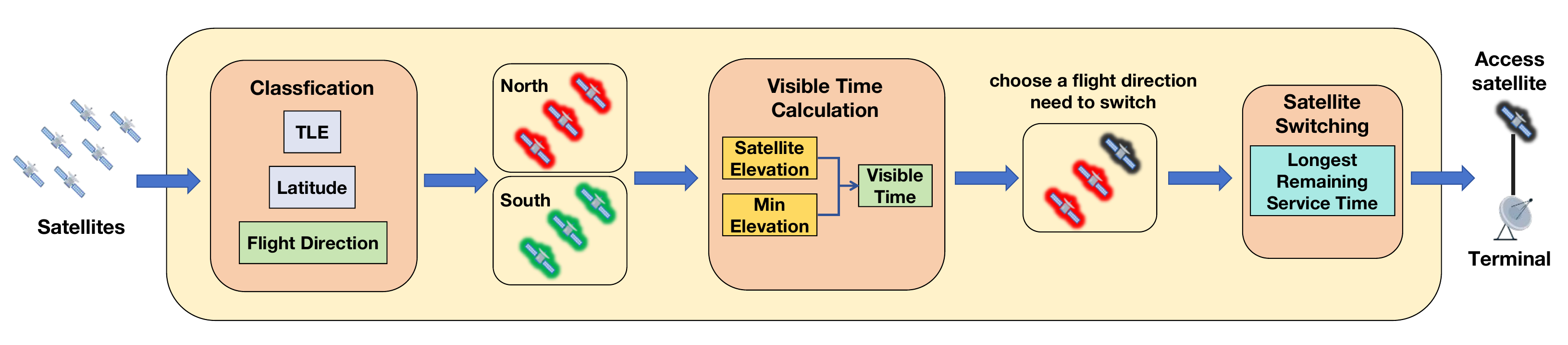}
    \caption{Overview of C-LRST. When GSL switches, we can get a target access satellite through C-LRST. First, there are a series of visible satellites over the terminal. Through Classification operation, they are classified according to their flight directions. The output of this operation is the north set and the south set of satellites. Second, choose a flight direction that needs to be switched, such as the north. Then, calculate the visible time of all satellites in this set. Finally, select the satellite with the longest remaining service time as the target access satellite of the terminal.}
    \label{fig:overview}
    \hfill
\end{figure*}

% modified 11.21
\section{Overview Design}
\label{sec:overview}
The overview design of C-LRST is shown in Fig.~\ref{fig:overview}. C-LRST has three main core components, \textit{Satellite Classification} is used to classify the visible satellites of the terminal according to their flight directions. \textit{Visible Time Calculation} is used to calculate the remaining service time of the visible satellites. Finally, the satellite with the longest remaining service time is selected as the access satellite through \textit{Satellite Switching}. Next, we will briefly introduce its core components.

\subsection{Satellite Classification}
The satellites above a terminal can be divided into north and south flight directions, respectively along the orbit up and down. When selecting the access satellite, the flight direction of the satellite will affect the hop number of the ISL~\cite{analysis_of_ISL}. Based on the TLE data of the satellite, the latitudes of the current time slot and the next time slot can be calculated, and the flight direction of the satellite can be inferred, and then make classification. The access satellites of the terminal's two GSLs come from the sets of two different flight directions of its visible satellites.

\subsection{Visible Time Calculation}
In order to determine the satellite with the longest remaining service time in the classified set, the visible time of the satellites to the terminal needs to be calculated. Calculate the position of the satellite relative to the terminal to obtain its actual elevation $e$, and check whether it is greater than $e_m$. Since the movement of satellites is regular, the expected time for the satellite to leave the visible range of the terminal can be obtained based on the above calculation. The difference between the expected time and the current time is the visible time of the satellite.

\subsection{Satellite Switching}
When GSL switches, the final satellite selection is made based on the determined flight direction set and the satellite with the longest service time determined in the set. Since the terminal maintains two GSLs, when the satellite connected to one of the GSLs is switched, we only need to first determine the set of access satellites that is different from the other, and then determine the satellite with the longest remaining service time, and finally complete the satellite selection. 

\section{Algorithm Design}
\label{sec:algorithm}
In order to support multi-access capabilities, we propose the Classification-based Longest Remaining Service Time (C-LRST) algorithm. In this section, we first present two propositions that support our algorithm. Second, we present our algorithm in detail. Finally, we introduce the execution flow of our algorithm through an example.

\begin{table}[!hpt]
    \caption{symbols of definitions}
    \label{tab:symbols}
    \centering
    \begin{tabular}{|c|l|}
    % \begin{tabularx}{\columnwidth}{lX}
    \hline
    Notation            & Defintion \\ 
    \hline
    
    $ N $              & Orbit number   \\
    $ M $              & \added{In-orbit satellite number}\\
    $ F $              & Factor \\
    $ \alpha $         & Orbit inclination \\
    \hline

    \added{$ O $}       & \added{The coverage area}   \\
    $ P $               & Any point in the area   \\
    $ D $               & Segment length    \\
    $ CL $              & Chord length   \\
    $ ST $              & Service time   \\    
    \hline

    $ T $               & Set of time slot \\
    $ ti $              & Time slot $i$\\
    $ S $               & \added{Set of satellites}\\
    $ s_i $             & Satellite $i$\\
    $ G $               & \added{Set of ground stations}\\
    \added{$ g_i $}     & \added{Ground station} $i$\\
    $ I $               & \added{Set of ISLs} \\
    $ L $               & \added{Set of GSLs} \\
    \added{$ l^{d}_{i} $}       & \added{GSL of $g_i$} \\
    $ C $               & Set of visible satellites \\
    $ e $               & Elevation angle \\
    $ e_m $             & Minimum elevation angle \\
    $ d $               & Flight direction of satellites \\
    \hline
    \end{tabular}
\end{table}

\subsection{Symbols and Definitions}
The key symbols used in this paper and their definitions are shown in TABLE ~\ref{tab:symbols}.
$ T $ is a set of time slots in which continuous time is evenly divided according to the time slot size, $T = \{t_0,t_1,\dots \} $. $ S $ is the satellite set, $S = \{s_0,s_1,\dots \} $. $ G $ is the set of ground stations, $G = \{g_0,g_1,\dots \} $. $ L $ is the GSL, $L = \{L_0,L_1,\dots \} $ and $L_i = \{l_i^0,l_i^1 \}$. $L_i$ record the satellites connected to $g_i$. $l_i^0$ is the GSL connected to the satellites flying north and $l_i^1$ is the GSL connected to the satellites flying south. $C$ is the set of visible satellites of $g_i$. $e_m$ is the minimum elevation angle. $d$ is the flight direction of satellites.$d$ can be north or south.  

\subsection{Theoretical Analysis}
\label{sec:theore}
To support our algorithm, we have the following two propositions.

First, we put forward Proposition~\ref{prp:1}, analyzing the satellite switching interval through mathematical derivation. Through Proposition~\ref{prp:1}, we finally obtain the average number of link switchings within a certain time range, which is used to prove the rationality of the result of our algorithm. 
Second, based on Proposition~\ref{prp:1}, we further put forward Proposition~\ref{prp:2}. By analysing that satellites in the north and south flight directions have the same service time distribution, we explain the rationality of our classification operation.

\begin{figure}[htb]
    \centering
    \includegraphics[width=0.8\linewidth]{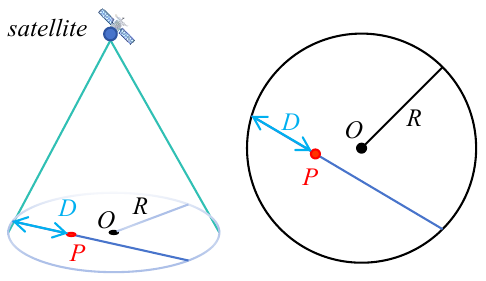}
    %\captionsetup{justification=centering} % 设置图标题居中
    \caption{Satellite service time calculation demonstration. The circle $O$ represents the coverage area of the satellite, $R$ is the radius of the circle, $P$ is any point within the circle, and $D$ represents the remaining service time along the satellite flight direction.}
    \label{fig:circle}
    \hfill
\end{figure}

\begin{Prp1}
  \normalfont
  The actual service time \added{$ST$} of the satellite is much shorter than the theoretical maximum service time \added{$ST_{max}$} of the satellite, and the number of switchings obtained by the C-LRST algorithm is reasonable.
\label{prp:1}
\end{Prp1}

% \textbf{Proposition 1.} The actual service time of the satellite is much shorter than the theoretical maximum service time of the satellite, and the number of switchings obtained by the C-LRST algorithm is reasonable.
\begin{Prf1}
   \normalfont
    We will analyse the service time and the number of switchings of the satellite with a specific beam range.
    
    The remaining service time of a terminal's access satellite is limited by the service time of a single satellite, which will have an impact on the GSL switching interval. We will calculate this service time below.
    
    As shown in Fig.~\ref{fig:circle}, the beam coverage of the satellite on the ground can be approximately regarded as a circle $O$, in which the service time $T$ to a terminal is related to the beam range, beam moving direction, and speed. Assume that the ground terminal appears with equal probability at any position within the satellite beam range. Although in actual situations, the satellite's beam will move in a fixed direction, in this analysis, considering the symmetry of the circle, we can still assume that the beam will move in any direction, so the position of the ground terminal in the beam can be expressed as any point $P$ on any chord on the circle. The remaining service time $ST$ of the satellite to $P$ is approximately proportional to the length of segment $D$ on the chord.
    
    Therefore, our problem can be transformed into calculating the length of $D$.
    
    Assume the chord length is $CL$. Since $P$ is uniformly distributed on $CL$ with equal probability, the probability density function of $D$ is $f(D) = \frac{1}{CL}$.We can get the length of $D$.
    \begin{equation}
        \int_0^{CL} f(D)l\, dl = \left. \frac{1}{CL} \cdot \frac{l^2}{2} \right|_0^{CL} = \frac{CL}{2}
    \end{equation}
    
    The problem is then transformed into calculating the length $CL$ of any chord, assuming that the distance from the chord to the center of the circle is $x$. Since the chord is chosen arbitrarily, $x$ is uniformly distributed within the range of $(0, R)$. The formula for the chord length is:
    \begin{equation}
        CL =2 \sqrt{R^2 - x^2}
    \end{equation}
    
    in which $x \in (0, R)$. The average chord length is:
    \begin{equation}
        \overline{CL} = \frac{1}{R} \int_0^R 2\sqrt{R^2 - x^2} \, dx = \frac{\pi \cdot R}{2}
    \end{equation}
    
    Therefore, the average length of $D$ is:
    \begin{equation}
        \overline{D}=\frac{ \overline{s} }{2}=\frac{\pi \cdot R}{4}
    \end{equation}
    
    Due to the direct proportional relationship, if the maximum service time $ST_{max}$ of the satellite is known, the average service time $\overline{ST}$ of the satellite can be calculated as:
    \begin{equation}
        \overline{ST} = \frac{\pi \cdot ST_{max}}{8}
    \end{equation}
    
    For routes with ISL, the access satellites at both ends may switch. Therefore, when the beams are continuously connected, for a certain time $t$, the average number of link switchings $n$ is:
    \begin{equation}
        n = \frac{2t}{ \overline{ST}}
    \end{equation}

    \added{From the above, we can conclude that our Proposition~\ref{prp:1}, actual service time is shorter than the theoretical maximum service time. For example, the $ST_{max}$ of a satellite is estimated to be roughly 250s (satellite at 550km altitude, elevation angle 25°). After calculation, the average service time $\overline{ST}$ of a single satellite is 98s. Within $t=1000s$, the average number of link switchings $n$ is about 20, which is similar to our subsequent experiment results.}
\end{Prf1}

Through the proof of Proposition~\ref{prp:1}, we can also obtain a following proposition.

\begin{Prp1}
  \normalfont
  Satellites in the north and south flight directions have the same service time distribution, so the classification selection operation is reasonable.
\label{prp:2}
\end{Prp1}

\begin{Prf1}
   \normalfont
    According to the proof process of Proposition~\ref{prp:1}, satellites have an average service time $\overline{ST}$ for terminals at any location within the beam. Therefore, a ground station with a certain position is uniformly random in the beam range of the visible satellites above. Therefore, whether the satellite is flying southward or northward, the service time of the ground station has the same service time distribution.

    Although the service times may vary for specific instances, overall, due to the regular construction of the constellation, satellites in both flight directions have the same service time expectations for ground stations.
\end{Prf1}

These two propositions will support our algorithm description and subsequent experiment results in Section~\ref{sec:experiments}, proving the correctness of the results and the rationality of the classification operation.

\begin{figure*}[htb]
    \centering
    \includegraphics[width=0.9\linewidth]{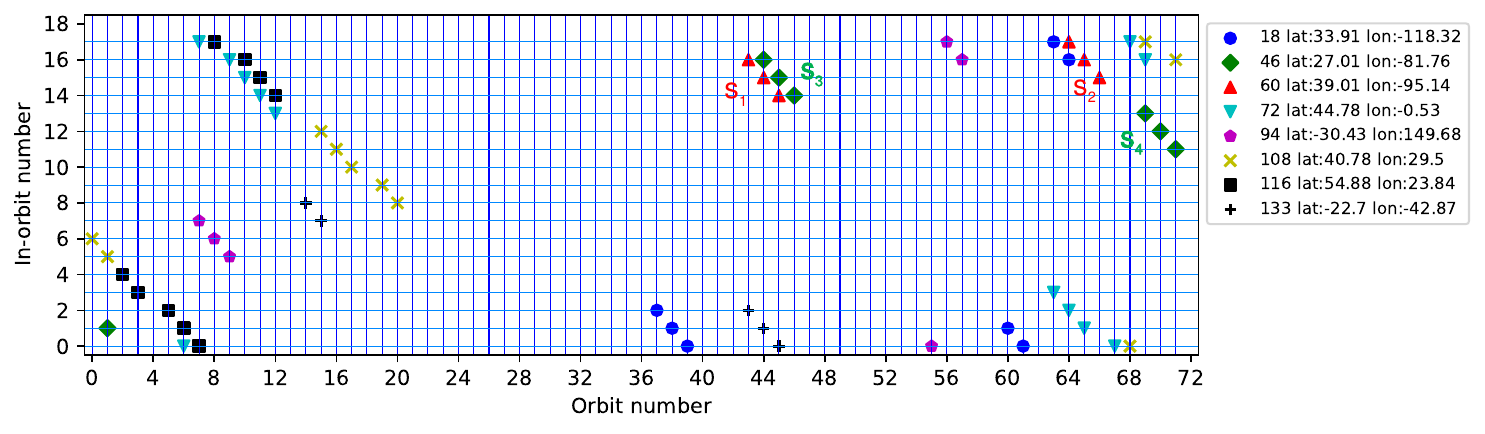}
    %\captionsetup{justification=centering} % 设置图标题居中
    \caption{\added{Visible satellites distribution of some ground stations.} Visible satellites above the ground station fly in two directions, forming two sets. \added{For example, t}he visible satellites of ground station 60 are labeled as two sets $s_1$ and $s_2$, and those of ground station 46 are labeled as two sets $s_3$ and $s_4$.}
    \label{fig:sat_visbile_of_gs}
    \hfill
\end{figure*}

\subsection{Algorithm Description}
\subsubsection{C-LRST}
The idea of the longest remaining service time is to constrain the terminal to select the satellite with the longest remaining service time during GSL switching, reducing the frequency of GSL switching and thereby minimizing network performance fluctuations.
Although this greedy strategy achieves its objective to a certain extent, the cost is unjustifiable. \added{We analyse the visible satellites distribution of ground stations}, as shown in Fig.~\ref{fig:sat_visbile_of_gs}. It is evident that there is a significant clustering phenomenon: visible satellites above the ground station fly in two directions, forming two sets. The visible satellites of ground station 60 are labeled as sets $s_1$ and $s_2$, and those of ground station 46 are labeled as sets $s_3$ and $s_4$.

Without a classification strategy, the satellite with the longest remaining service time chosen by ground station 60 may come from $s_1$ or $s_2$, and station 46 may come from $s_3$ or $s_4$. This leads to the following problem: if both ground stations ultimately select access satellites from $s_1$ and $s_3$, the maximum number of ISL hops will not exceed 5. However, if the selected access satellites come from $s_1$ and $s_4$, the minimum number of ISL hops will not be less than 25. The RTT ratio in both cases will be over 5x. It's not worth sacrificing such a large RTT in exchange for longer service time.

C-LRST algorithm consists of three parts: core, initialization, and handover. The function of the Core part Algorithm~\ref{algo:C-LRST-part1} is to obtain the longest remaining service time satellite corresponding to the flight direction at time $t$, when the ground terminal and satellite flight direction are specified. $C^0$ and $C^1$ respectively represent the set of visible satellites flying northward and southward of the ground terminal $g_i$. Lines 3-10 are used to calculate $C^0$ and $C^1$. Lines 11-14 determine the set of candidate satellites based on the input flight direction. Lines 15-19 obtain the satellite with the longest service time in the candidate satellite set. The computational complexity of C-LRST core part is $O(n)$.

% The function of the initialization part Algorithm~\ref{algo:C-LRST-part2} is to determine the two satellites to which each ground terminal needs to connect at the initial moment. To obtain $L_{G,0}$, Algorithm~\ref{algo:C-LRST-part1} needs to be called twice for each ground terminal at time $0$.
\added{The function of the initialization part Algorithm~\ref{algo:C-LRST-part2} is to determine the two access satellites for each ground terminal at the initial moment. To obtain $L_{G,0}$, Algorithm~\ref{algo:C-LRST-part1} needs to be called twice for each ground terminal at time 0.}

When a satellite, which is connected to a ground terminal through a GSL, flies out of the visible range, the GSL needs to be updated. Algorithm~\ref{algo:C-LRST-part3} is used to check whether any ground terminal needs to update GSL in each time slot. \added{If an access satellite at time $t$ flies out of the visible range at time $t+1$, Algorithm~\ref{algo:C-LRST-part1} is called to obtain a new access satellite with the same flight direction.}

\begin{algorithm}[t]
\caption{C-LRST}
\label{algo:C-LRST-part1}
\KwIn{$g_i$, $S$, $d$, $e_{m}$, $t$} 
\KwOut{$c$}
$C^0 = \emptyset$ \\
$C^1 = \emptyset$ \\
\For{$s_j \in S$}{
    calculate $e_{i,j}$ as the elevation angle of $s_j$ relative to $g_i$ at time $t$\\
    \If{$e_{i,j} \added{\geq} e_m$}{
        calculate $d_j$ as the flight direction of $s_j$ at time $t$\\
        \If{$d_j$ is north}{
            $C^0=C^0 \cup \{s_j\}$
        }
        \If{$d_j$ is south}{
            $C^1=C^1 \cup \{s_j\}$
        } 
    }
}
\If{\added{$d$} is north}{
        $C = C^0$
}
    \Else{
        $C = C^1$
}
$r = 0$, $c = 0$ \\
\For{$s_j \in C$}{
     calculate $r_{i,j,t}$ as the remaining service time of $s_j$ relative to $g_i$ at time $t$\\
     \If{$r_{i,j,t}\geq r$}{
        $r = r_{i,j,t}$, $c = s_j$\\
     } 
}
\end{algorithm}

\subsubsection{Proof of correctness}
We illustrate the correctness of our classification satellite selection algorithm through induction.
First, for a terminal, consider that there is only one visible satellite above it, then it will be chosen as the access satellite.
Second, we consider that there are two visible satellites. If the two satellites belong to the same flight direction, the one with the longest remaining service time will be selected as the access satellite. If the two satellites belong to different flight directions, they will both be chosen to establish GSLs.
Third, we consider that there are multiple visible satellites, which is a general case. According to the C-LRST algorithm, visible satellites are classified, and satellites with the longest remaining service time are selected for connection.
According to Proposition~\ref{prp:2} in Section~\ref{sec:theore}, satellites flying in the north and south flight directions have the same service time distribution. 
Therefore, it is reasonable to establish GSLs with satellites that have the longest remaining service time in both north and south flight directions by classifying the visible satellites.

\begin{figure*}[htb]
    \centering
    \includegraphics[width=0.9\linewidth]{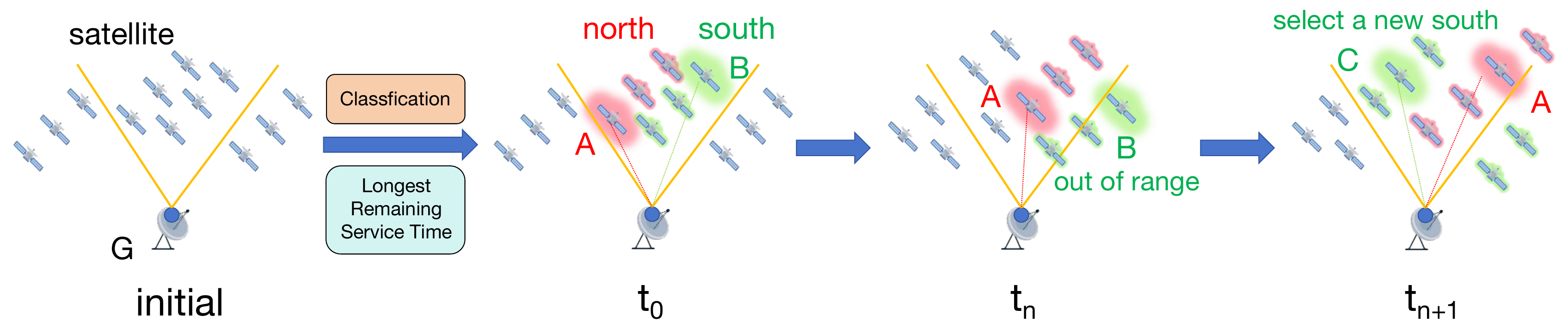}
    % \captionsetup{justification=centering} % 设置图标题居中
    \caption{Execution process of C-LRST Algorithm. This figure shows the initialization operation of ground station $G$ and the satellite switching operation at time $t_{n+1}$. Initially, there are some satellites above ground station $G$, among which the satellites located within the elevation angle range are visible satellites of $G$. At time $t_0$, the visible satellites are classified into north and south sets according to their flight directions. The satellites with the longest service time are $A$ and $B$ respectively, which are selected as access satellites. Then, at time $t_n$, $A$ is still within the visible range of $G$, while $B$ leaves the range. Therefore, a new satellite $C$ will be selected from the south set at the next moment $t_{n+1} $.}
    \label{fig:example}
    \hfill
\end{figure*}

\begin{algorithm}[t]
\caption{Initialization}
\label{algo:C-LRST-part2}
\KwIn{$G$, $S$, $e_m$} 
\KwOut{$L_{G,0}$}
$L_{G,0} = \emptyset $ \\
\For{$g_i \in G$}{
    $l_{i,-1}^0 = -1$, $l_{i,-1}^1 = -1$ \\
    calculate $c_0$ and $c_1$ as the candidate satellites flying  north or south at time 0 by algorithm \ref{algo:C-LRST-part1} \\
    $l_{i,0}^0 = c_0$, $l_{i,0}^1 = c_1$ \\
    $L_{G,0} =  L_{G,0} \cup \{ l_{i,0}^0, l_{i,0}^1\} $
}
\end{algorithm}

\begin{algorithm}[t]
\caption{Handover}
\label{algo:C-LRST-part3}
\KwIn{$G$, $S$, $e_m$,$L_{G,t}$} 
\KwOut{$L_{G,t+1}$}
$L_{G,t+1} = \emptyset$ \\
\For{$g_i \in G$}{
    \If{$l_{i,t}^0$ is visible at time $t+1$}{
        $l_{i,t+1}^0 = l_{i,t}^0$
    }
    \Else{
        calculate $l_{i,t+1}^0$ as the candidate satellite flying north at time $t+1$ by algorithm \ref{algo:C-LRST-part1}
    }
    \If{$l_{i,t}^1$ is visible at time $t+1$}{
        $l_{i,t+1}^1 = l_{i,t}^1$
    }
    \Else{
        calculate $l_{i,t+1}^1$ as the candidate satellite flying south at time $t+1$ by algorithm \ref{algo:C-LRST-part1}
    }
    $L_{G,t+1} =  L_{G,t+1} \cup \{ l_{i,t+1}^0, l_{i,t+1}^1\} $
}
\end{algorithm}

\subsection{Example Analysis}
Next, we will describe the execution process of our algorithm through an example.

As shown in Fig.~\ref{fig:example}, there are a series of satellites above ground station $G$, among which the satellites located within the elevation angle range are visible satellites of $G$. At time $t_0$, the initialization operation is performed. The visible satellites are classified according to their flight directions, and the satellites with the longest service time are selected as the two access satellites of the ground station. The satellite flying northward selected by ground station $G$ is $A$, and the satellite flying southward is $B$. Assume that at time $t_n$, $A$ is still within the visible range of ground station $G$, while southward access satellite $B$ leaves the range. We need to select a new satellite $C$ with the longest service time from the southward set of visible satellites at the next moment $t_{n+1} $ as the new access satellite.

\section{Experiments}
\label{sec:experiments}
% haohao: 7.7
% \subsection{Environment}}
\subsection{\added{Experiment Setup}}
We use the open-source Hypatia simulation platform developed by Kassing et al.~\cite{hypatia}, which provides a LEO satellite network simulation framework. This platform allows for the pre-computation of network states over time and implements packet-level simulations using NS-3.

% haohao: 7.7
% \subsection{Dataset and Parameters}
In our experiments, we choose Starlink \textit{shell 1} for study. In addition, we select 165 ground stations registered by Starlink around the world as ground nodes~\cite{starlink.sx}. To ensure the adaptability of the algorithm to real satellite constellation scenarios, we conduct experiments using the actual structure of \textit{shell 1} derived from TLE data published by Starlink~\cite{liu2022starlink}. TABLE~\ref{tab:550_km_args} shows the detailed configurations.

% haohao: 7.7
% \subsection{Performance}
\added{Our experiments focus on the Longest Remaining Service Time algorithm that only maintains one GSL (LRST-1), the Longest Remaining Service Time algorithm that maintains two GSLs (LRST-2), the Coordinated Satellite-Ground Interconnecting algorithm mentioned in Zhang's paper (CSGI), and our proposed Classification-based Longest Remaining Service Time algorithm (C-LRST). The above algorithms are compared and analysed mainly based on the performance metrics of Ping, GSL switching interval, throughput, and TCP RTT.}

% haohao 7.7
% 补充算法的实现和模拟设置
% Hypatia通过读取卫星和地面站节点的配置信息，可以构建一定仿真时间内的动态拓扑，确定节点之间的路由连通状态，这是通过Python实现的。进一步地，Hypatia结合NS3构建具备网络功能的虚拟节点，从而实现对星座和地面站的模拟。
% 上述提到的算法包括我们提出的C-LRST，是不同的星地互联方案，对应在仿真模拟中会产生不同的动态拓扑和路由。因此，这些算法的实现是通过修改Hypatia的拓扑和路由构建方法实现的。我们在网上公开了我们的实现~\cite{TODO}。
\added{Hypatia can construct a dynamic topology and determine the routing connectivity status between nodes within a certain simulation time by reading the configuration information of the satellites and ground stations. This is implemented by Python. Furthermore, Hypatia combines NS3 to build virtual nodes with network functions, thereby simulating constellations and ground stations.
The algorithms mentioned above are different satellite-to-ground interconnection schemes that will produce different dynamic topologies and routings in the simulation. Therefore, these algorithms is implemented by modifying the topology and routing construction methods of Hypatia. We have made our implementation public online~\cite{LMCN-SGI}.}

% haohao: TODO 感觉这句话可以丢到后面
% \added{Based on Hypatia, we establish end-to-end TCP links between ground stations and conduct $1000s$ simulation experiments. In our experiment, t}he NS3 time granularity is $1ns$, the route update interval is $100ms$, and the link transmission rate is set to $10Mbit/s$. We use TCP Hybla as the congestion control algorithm. This algorithm is mainly used to solve the problem of low TCP throughput in long link scenarios, so it is often used in TCP applications of satellite networks~\cite{hybla}. The simulation experiment parameters are shown in TABLE~\ref{tab:sim_args}.

\begin{table}[!hpt]
    \caption{Starlink shell 1 parameter configuration}
    \label{tab:550_km_args}
    \centering
    \begin{tabular}{|c|l|}
    % \begin{tabularx}{\columnwidth}{lX}
    \hline
    \rule[-6pt]{0mm}{18pt}    Parameters  & Value \\ 
    \hline
    \rule[-6pt]{0mm}{18pt}    Orbit altitude & $550$ km \\ 
    \hline
    \rule[-6pt]{0mm}{18pt}    \added{Orbit number ($N$)} & 72 \\ 
    \hline
    \rule[-6pt]{0mm}{18pt}    \added{In-orbit satellite number ($M$)} & 18\\ 
    \hline
    \rule[-6pt]{0mm}{18pt}    Factor ($F$) & 45 \\ 
    \hline
    \rule[-6pt]{0mm}{18pt}    Orbit inclination ($\alpha$) & 53° \\ 
    \hline
    \end{tabular}
\end{table}

\subsection{The impact of frequent switching}
% 首先在这一节中，我们通过对比ISL mode 和 Hybird mode的性能差异来研究频繁的GSL切换的影响，从而说明ISL mode的优势。
\added{As mentioned in Section~\ref{sec:background-info}, we first study the impact of frequent GSL switching in this section by comparing the performance differences between ISL mode and Hybrid mode.}
% To explore the impact of frequent GSL switching on satellite networks, we compare the network performance differences between ISL mode and Hybrid mode. 
% The ISL mode means that the satellite only connects to the satellite through GSL at the ground stations at the source and destination, and the intermediate route only passes through the ISLs. The Hybrid mode adds ground nodes to the network topology, mixes the traditional transparent forwarding mode with the ISL mode, and finds the global shortest path.

\begin{figure}[htb]
    \centering
    \subfigure[Average Ping RTT]{
        \includegraphics[width=0.46\linewidth]{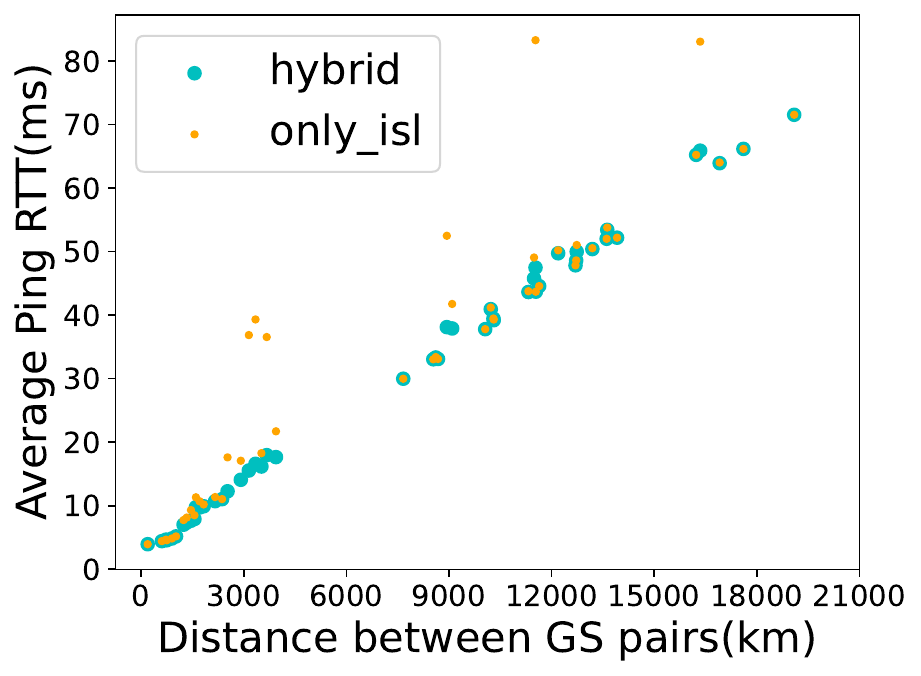}
        \label{subfig:average_ping_vs_distance}
    }
    \subfigure[CDF of Ping RTT]{
        \includegraphics[width=0.45\linewidth]{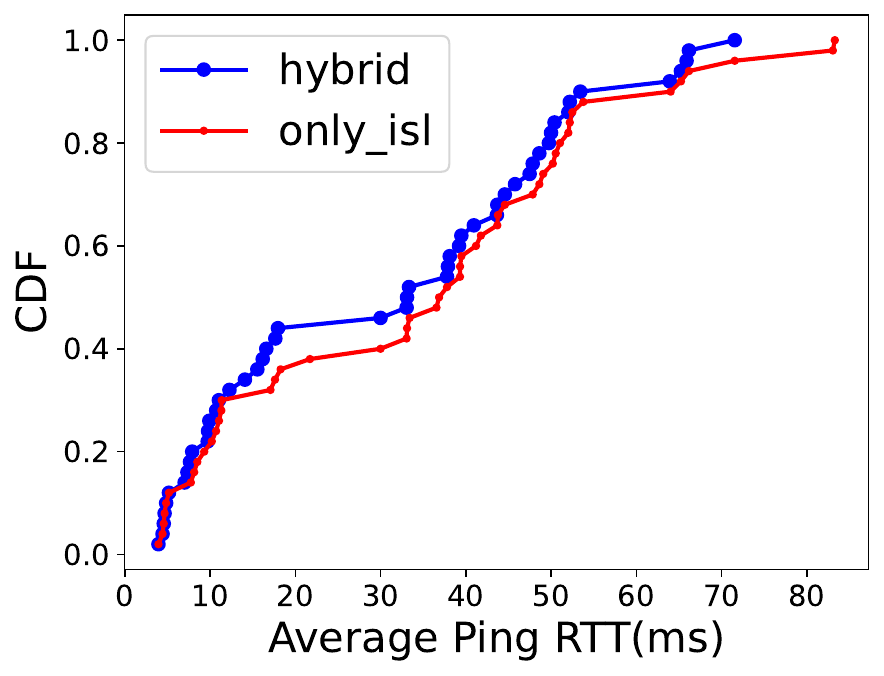}
        \label{subfig:cdf_of_average_ping}
    }
    \caption{\added{Average Ping RTT over distance between ground station pairs. In most cases, Hybrid mode and ISL mode have similar ping values. Only in a few pairs, Hybrid mode has shorter paths and smaller ping values.}}
    % 在大多数情况下，Hybrid mode和ISL mode有相近的ping value。只有少数的pair，Hybrid mode会有更短的路径选择，有较小的ping value。
    % \caption{Average Ping RTT over distance between ground station pairs. The cumulative distribution of the Ping RTT in ISL mode has only slight differences compared with Hybrid mode. This difference is mainly due to the poor path selection in ISL mode between a few ground station pairs.}
    \label{fig:combined_fig}
\end{figure}

\begin{figure}[htb]
    \centering
    \subfigure[Average Rate]{
        \includegraphics[width=0.46\linewidth]{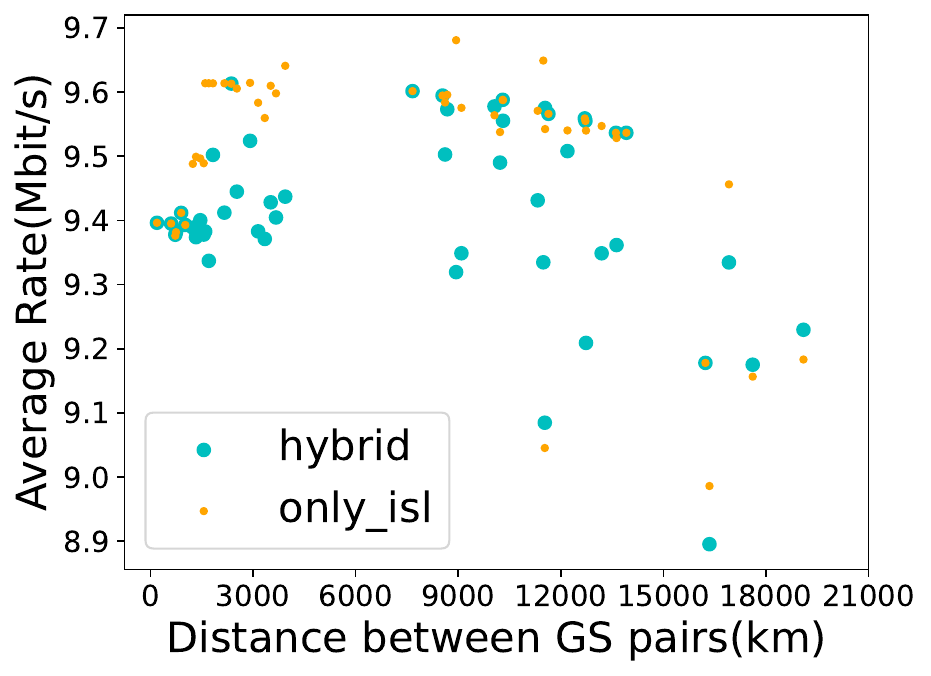}
        \label{subfig:average_rate_vs_distance}
    }
    \subfigure[CDF of Rate]{
        \includegraphics[width=0.45\linewidth]{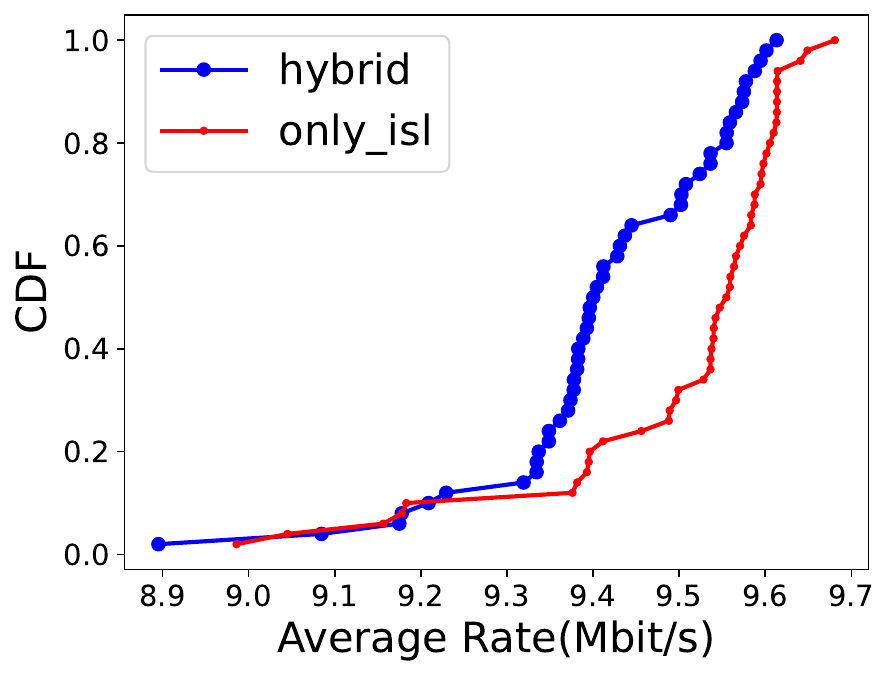}
        \label{subfig:cdf_of_average_rate}
    }
    \caption{Average Rate over distance between ground station pairs. In most paths, Hybrid mode exhibits poorer rate performance. This means that compared to Hybrid mode, ISL mode can achieve higher throughput.}
    \label{fig:cdf}
\end{figure}

\added{Fig.~\ref{subfig:average_ping_vs_distance} shows how the ping value changes with the distance between ground station pairs under the shortest path algorithm. This result verifies our hypothesis. During the selection of the shortest paths, for most ground station pairs, the two modes choose similar routes. For a few pairs, the paths chosen by the Hybrid mode exhibit lower ping values than the paths selected by the ISL mode. From the CDF of the average ping value shown in Fig.~\ref{subfig:cdf_of_average_ping}, we can also see that the overall distribution of ping value in ISL mode is slightly higher than that in Hybrid mode.} 
However, as can be seen from the results of Fig.~\ref{subfig:average_rate_vs_distance} and Fig.~\ref{subfig:cdf_of_average_rate}, shorter paths do not bring higher throughput, and the throughput in Hybrid mode is generally lower than that in ISL mode.
\begin{figure}[!htp]
    \centering
    \subfigure[\added{Hybrid Mode Rate}]{
        \includegraphics[width=0.46\linewidth]{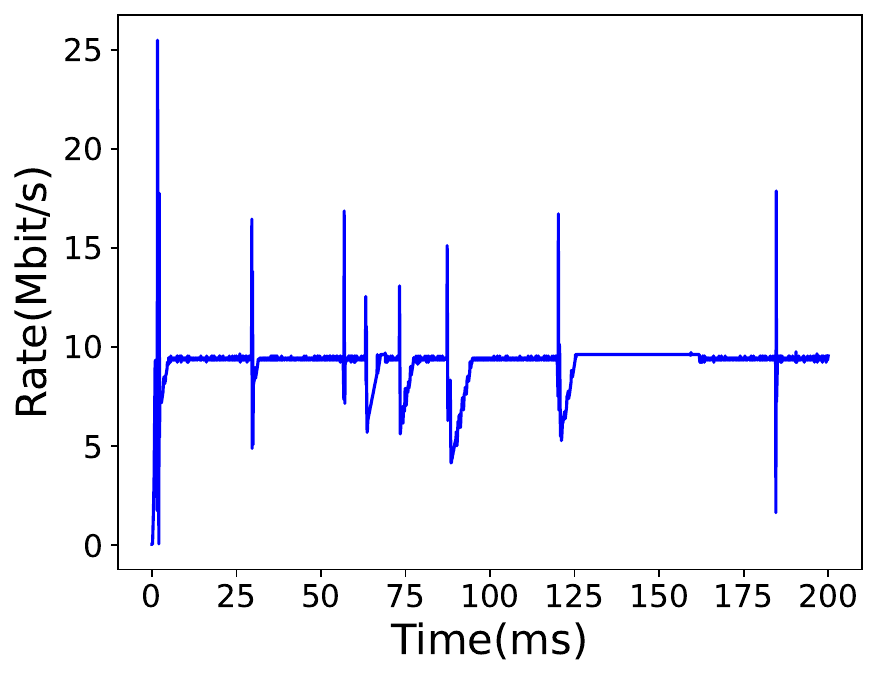}
        \label{subfig:hybrid_rate}
    }
    \subfigure[\added{ISL Mode Rate}]{
        \includegraphics[width=0.46\linewidth]{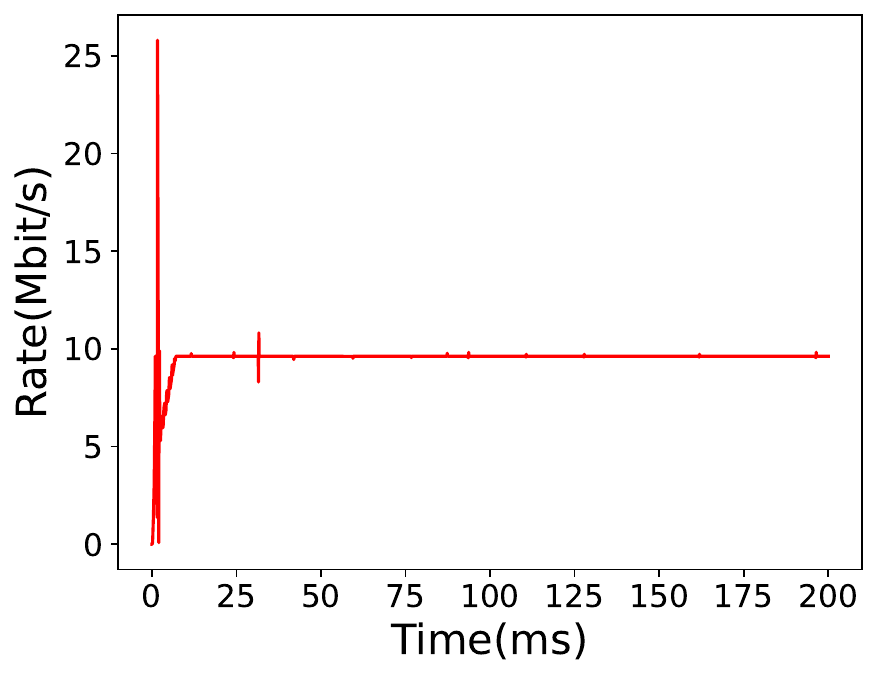}
        \label{subfig:only_isl_rate}
    }
    \caption{\added{Hybrid mode rate and ISL mode rate of a specific pair. This shows the reason for the different average rates above. More GSLs in Hybrid mode result in more switchings, which affect link stability and throughput.}}
    % \caption{Rate of Hybrid and only ISL. On some paths, although the overall rate performance of Hybrid mode can be close to that of ISL mode, the more frequent GSL switching in Hybrid mode leads to drastic fluctuations in the rate.}
    \label{fig:rate_of_hybrid_and_only_isl}
\end{figure}

\added{As shown in Fig.~\ref{fig:rate_of_hybrid_and_only_isl}, which describes the rate of a specific pair of ground stations. Between the same ground station pair, Hybrid mode switches much more frequently than ISL mode.} 
Multiple switchings occur within the $65-100ms$ period, which significantly reduces the throughput. The decrease in throughput is due to packet loss caused by frequent path changes. 
\added{However, in reality, in addition to the impact of path changes on throughput and stability, GSL switching itself also has a certain overhead~\cite{satcp_2023},}
% However, in reality, aside from the impact of path changes, GSL switching itself also has a certain overhead~\cite{satcp_2023}, 
but due to platform limitations, this aspect is not reflected in our study. Consequently, in practical applications, frequent GSL \added{switchings result in} worse performance.

\subsection{End-to-end path analysis}
% haohao 7.7
% 在本节中，我们首先关注一组特定的地面站节点对之间的通信，来详细地了解几种算法的性能，对各种指标的现象和原因进行分析。之后，我们研究了地面站节点之间的通信随距离的变化情况，从而对几种算法有一个更全面的理解。
\added{In this section, we first focus on the communication between a specific ground station pair to understand the performance of the four algorithms in detail and analyse the phenomena and causes of various indicators. Next, we study how communication between ground station nodes varies with distance to have a more comprehensive understanding of these algorithms.}

\begin{table}[!hpt]
    \caption{List of simulation experiment parameters}
    \label{tab:sim_args}
    \centering
    \begin{tabular}{|c|l|}
    % \begin{tabularx}{\columnwidth}{lX}
    \hline
    \rule[-6pt]{0mm}{18pt} Parameters  & Value \\
    \hline
    \rule[-6pt]{0mm}{18pt} Simulation time & 1000 s  \\
    \hline
    \rule[-6pt]{0mm}{18pt}  Simulation time granularity & 1 ns  \\
    \hline
    \rule[-6pt]{0mm}{18pt} Route update interval   & 100 ms   \\
    \hline
    \rule[-6pt]{0mm}{18pt} Transmission rate of ISL and GSL    & 10 Mbit/s \\
    \hline
    \rule[-6pt]{0mm}{18pt} TCP congestion control algorithm    & TCP Hybla \\
    % Initial cwnd    & 10 packets  \\
    % Buffer size of ISL and GSL  & 0.655 MB  \\
    % Segment size    & 1380 byte \\
    \hline
    \end{tabular}
\end{table}

\added{Based on Hypatia, we establish end-to-end TCP links between ground stations and conduct $1000s$ simulation experiments. In our experiment, the NS3 time granularity is $1ns$, the route update interval is $100ms$, and the link transmission rate is set to $10Mbit/s$. We use TCP Hybla as the congestion control algorithm. This algorithm is mainly used to solve the problem of low TCP throughput in long link scenarios, so it is often used in TCP applications of satellite networks~\cite{hybla}. The simulation experiment parameters are shown in TABLE~\ref{tab:sim_args}.}

As shown in Fig.~\ref{fig:ptp_location}, the ground stations used in the experiment are located in Itaboraí (Brazil) and Kaunas (Lithuania). The straight-line distance between the two ground stations is approximately $10,000km$. 

\subsubsection{Ping}
\begin{figure}[htb]
    \centering
    \subfigure[Ground station pair from Itaboraí (Brazil) to Kaunas (Lithuania).]{
        \includegraphics[width=0.6\linewidth]{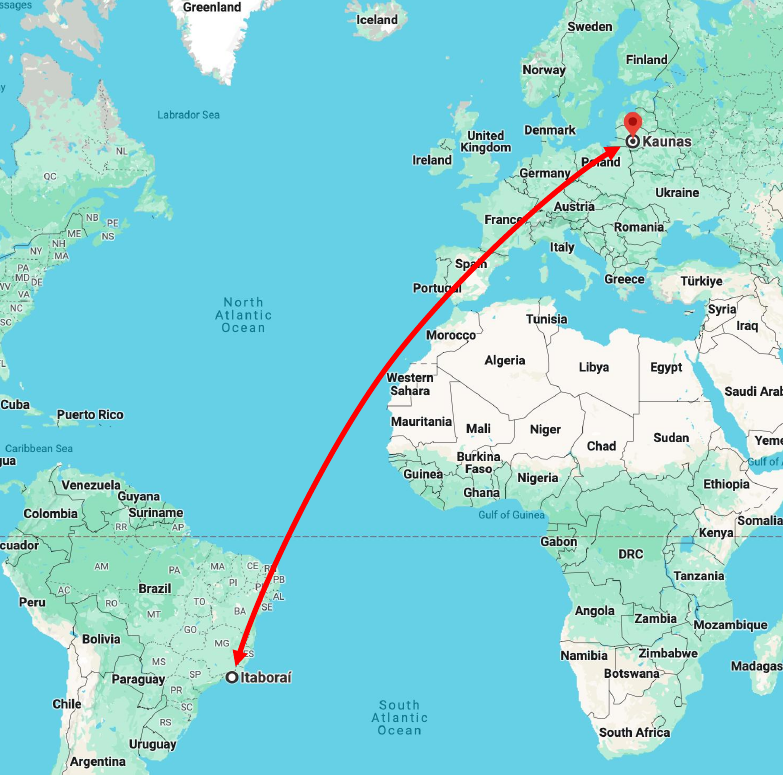}
        \label{fig:ptp_location}
    }
    \hfill
    \subfigure[Ping RTT. Compared to other algorithms, C-LRST has a lower Ping RTT. \added{The fluctuations are due to the inevitable GSL switchings.}]{
        \includegraphics[width=0.46\linewidth]{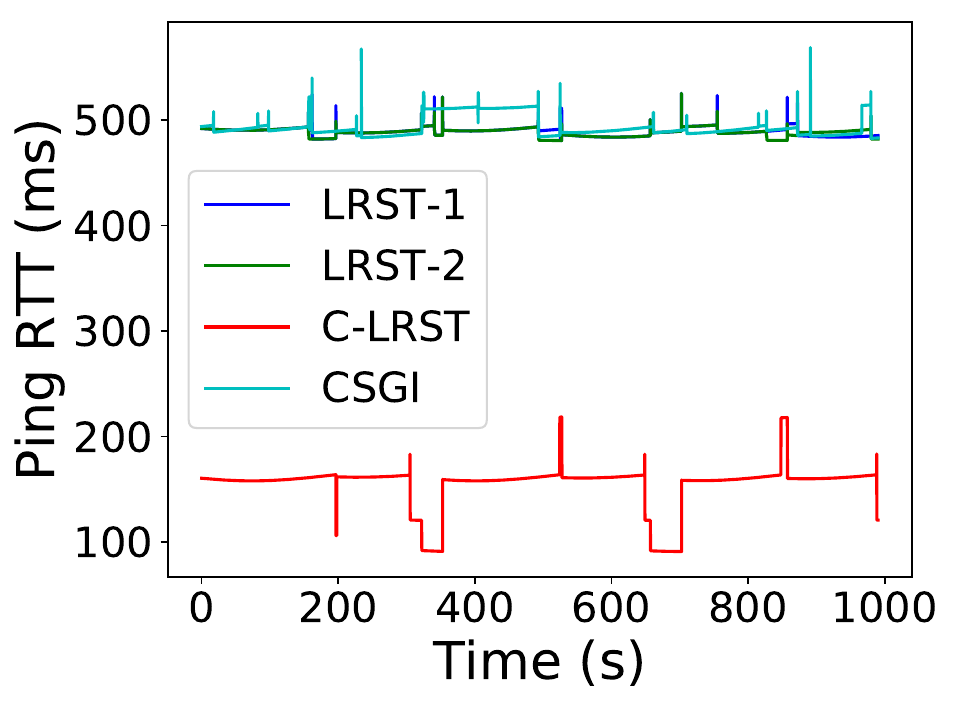}
        \label{fig:ping_1429_to_1412}
    }
    \hfill
    \subfigure[\added{Switching interval. The results show that C-LRST has relatively larger switching intervals and lower switching frequency.} 
    %On this path, C-LRST has 14 switchings, whereas CSGI has 23.
    ]{
        \includegraphics[width=0.46\linewidth]{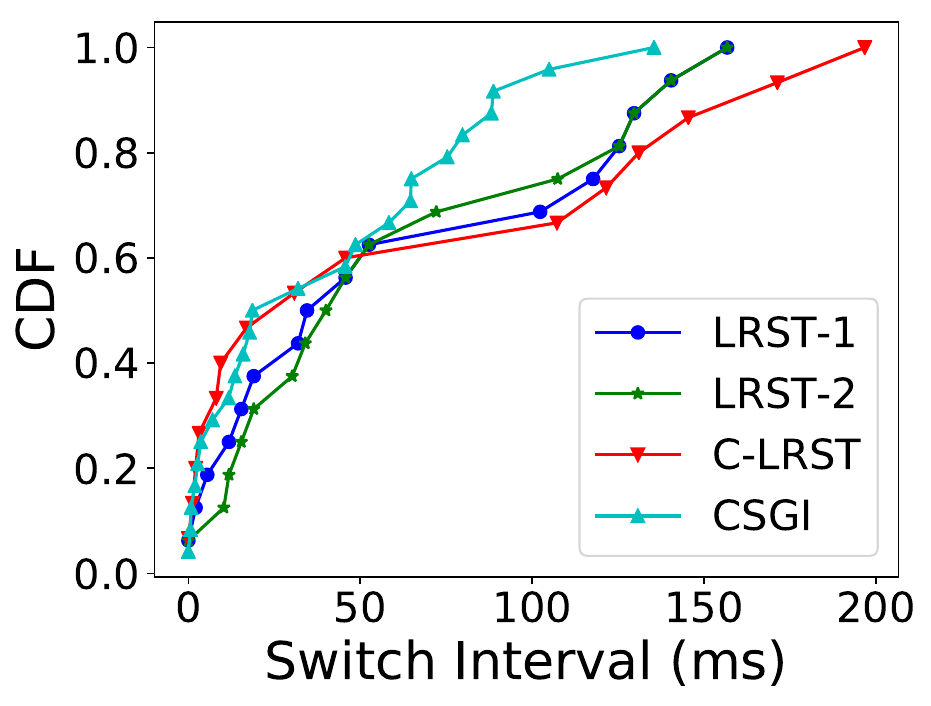}
        \label{fig:interval_1429_to_1412}
    }
    \caption{\added{Ground station pair, Ping RTT, and GSL switching interval.}}
    \label{fig:ping RTT and ptp switch interval}
\end{figure}

Fig.~\ref{fig:ping_1429_to_1412} shows the ping value during the $1000s$ simulation under different algorithms. The ping values of LRST-1, LRST-2, and CSGI are all high, all around $500ms$, and they are unstable. The ping value of C-LRST is relatively low, mostly staying around $160ms$, with only a few jitters.

Based on the previously mentioned idea of classifying visible satellites, we can explain the ping values of the above algorithms: the ping value of LRST-1 is much higher than that of the other algorithms. This is because the access satellites selected by the two ground stations are located in two far-together categories. For the CSGI algorithm, although the path length is optimized from a global perspective, it is limited to connecting only one satellite and faces the same problems as LRST-1. Although LRST-2 maintains two GSLs, the ping value has almost no improvement compared with LRST-1. This is because the two satellites selected by each ground station during this period are in the same category, and the two categories are far apart. The C-LRST we proposed further classifies visible satellites on the basis of maintaining two GSLs, effectively avoiding the above-mentioned problems, and therefore has better performance.

\begin{figure*}[htb]
  \centering
    \includegraphics[width=0.7\linewidth]{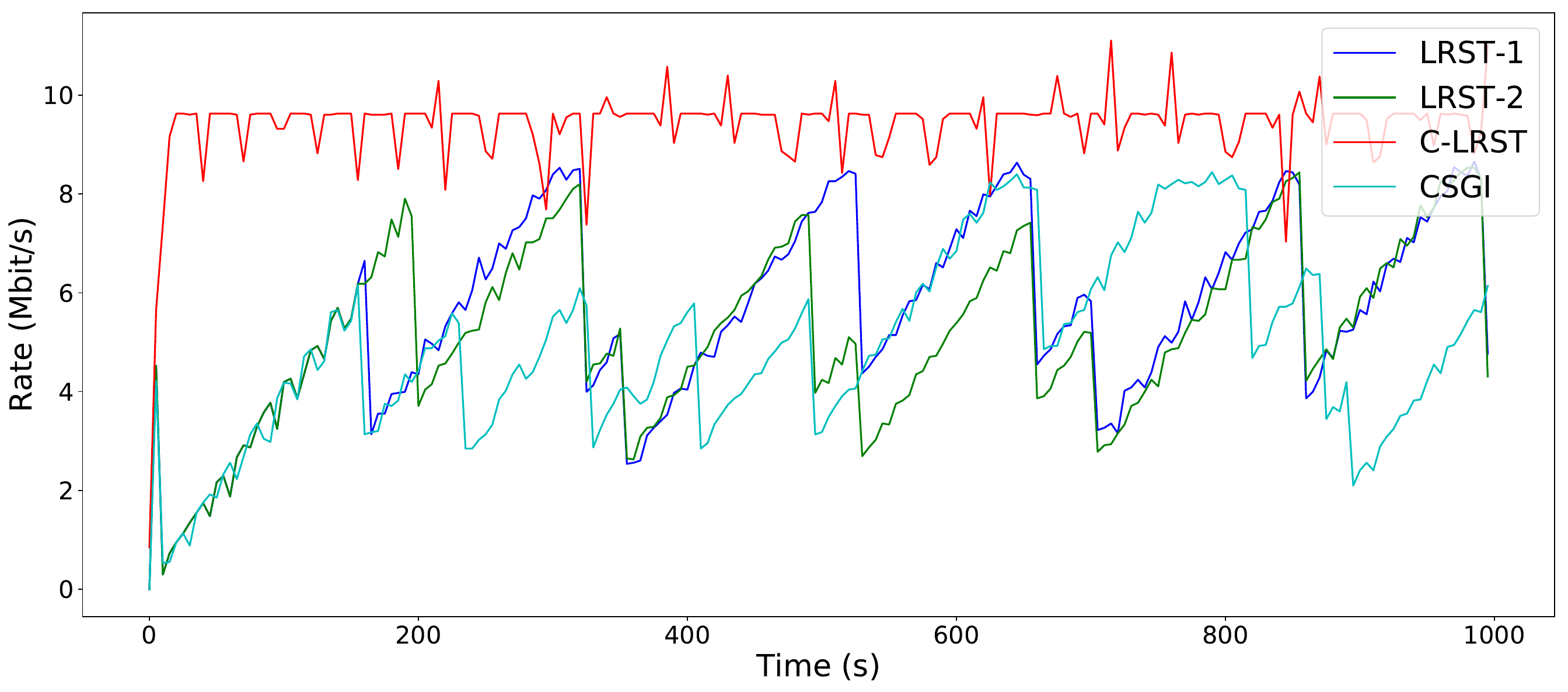}
    \caption{\added{Rate and Throughput}. Over the entire simulation time, C-LRST has significantly stable and higher rate than other algorithms, which means that C-LRST has better throughput performance. The low rate of CSGI and other algorithms is due to the inefficient congestion handling caused by long latency, and the high GSL switching frequency.}
    \label{fig:rate_1429_to_1412}
\end{figure*}

\begin{figure}[htb]
    \centering
    \includegraphics[width=\linewidth]{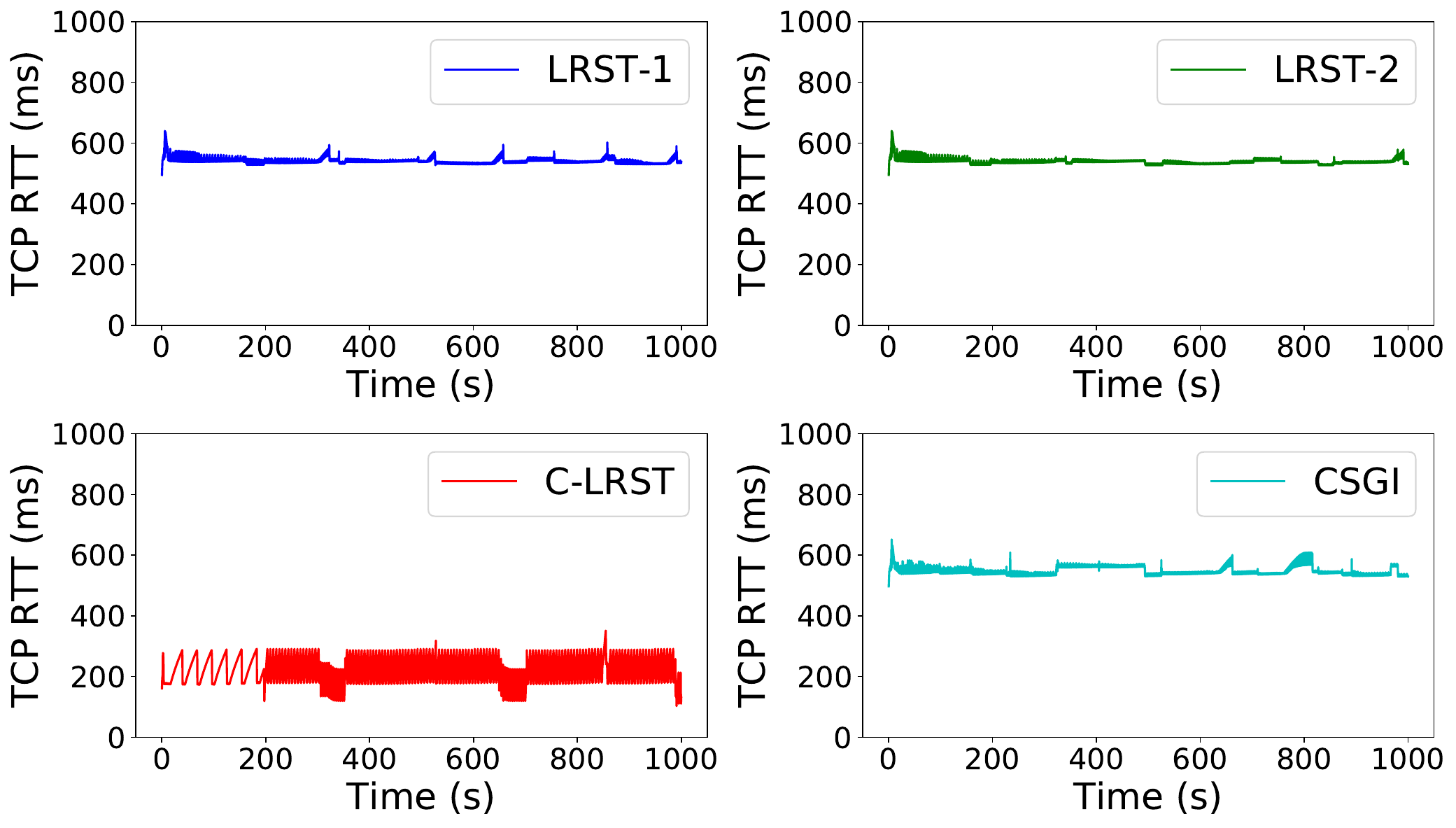}
    % \subfigure[TCP RTT of each algorithms.]{
    %     \includegraphics[width=\linewidth]{figure/rtt_1429_to_1412.pdf}
    % }
    % \subfigure[Average TCP RTT of each algorithms.]{
    %     \includegraphics[width=0.7\linewidth]{figure/bar_1429_to_1412.pdf}
    % }
    \caption{TCP RTT. Compared to Ping RTT, TCP RTT also includes additional time such as packet processing, so it is slightly higher. However, the overall variation is consistent, and C-LRST has a lower RTT. The fluctuations are due to the characteristics of the TCP Hybla congestion control algorithm and have only a negligible effect.}
    \label{fig:rtt_1429_to_1412}
\end{figure}

% modified 11.20
\subsubsection{GSL switching interval}
% Fig.~\ref{fig:1429_to_1412_interval} shows the switching time interval of the access satellites selected by the two ground stations in the 1000s simulation. As long as the access satellites of any ground station changes, it is regarded as a switching. We infer that access satellite changes are the main cause of ping jitter, and Fig.~\ref{fig:1429_to_1412_interval} verifies this conjecture. Among them, the overall switching frequency of LRST-1, LRST-2, and C-LRST is relatively low, making full use of the GSL that can be used for a longer time, while the CSGI switching frequency is slightly higher.
Fig.~\ref{fig:interval_1429_to_1412} shows the time interval between each switching of the ground station's access satellite in a $1000s$ simulation. As long as the access satellite of either of the two ground stations changes, it is considered a switching. The switching of access satellite is the main cause of ping jitter.

As can be seen from Fig.~\ref{fig:interval_1429_to_1412}, for LRST-1 and LRST-2, 14 switchings occur in the entire simulation process, C-LRST is 15, while CSGI is the most, up to 23. In addition, LRST-1, LRST-2, and C-LRST are at the lower right, which indicates that there is a longer time interval between two switchings, and the overall switching frequency is relatively low, making full use of the GSL that can be used for a longer time. The CSGI curve is near the upper left, the switching interval is short, and the switching frequency is slightly higher. Fig.~\ref{fig:interval_1429_to_1412} further explains that in Fig.~\ref{fig:ping_1429_to_1412}, CSGI has more ping jitters.

For CSGI, due to the limitation of connecting to only one satellite, even if it considers the overall situation, it cannot optimize the service duration and reduce the switching frequency for all ground stations. For example, in the experiment on the node pairs we selected above, the algorithm did not perform well. The C-LRST algorithm we proposed contains the idea of the longest remaining service time, retaining the advantage of lower switching frequency.

\subsubsection{Throughput}
Fig.~\ref{fig:rate_1429_to_1412} shows the throughput changes at the sender for different algorithms. The throughputs of the three algorithms LRST-1, LRST-2, and CSGI are low and unstable. For the C-LRST algorithm, the throughput can be stabilized at approximately $10Mbit/s$ relatively quickly, jittering occurs only at a few moments, and quickly recovers to $10Mbit/s$.

One reason for low throughput is long delay. First, the slow start time is delayed due to the longer RTT. Second, a longer RTT means a longer congestion control period. For example, congestion that occurs at the receiving end takes a longer time to be perceived by the sending end, which makes congestion be processed more slowly and has a greater impact on throughput. In addition, due to the expansion of the scope of congestion, TCP Hybla will have fewer opportunities for fast retransmission and will instead enter a longer congestion avoidance state~\cite{hybla}.

Another reason for low throughput is the high frequencies of link switching. As Fig.~\ref{fig:rate_1429_to_1412} shows, although CSGI imposes constraints on the minimum service time of access satellites, it is still unavoidable that multiple handovers occur in a short period. For example, around $160s$ and $660s$, multiple switchings bring about a sharp decrease in throughput. 
At $405s$, because the access satellites of two terminals are switched from (792, 56) to (827, 73), the ISL increases by 14 hops, which also causes a sharp decrease in throughput.

\subsubsection{\added{TCP RTT}}
\added{Fig.~\ref{fig:rtt_1429_to_1412} shows the performance of the TCP RTT using TCP Hybla congestion control algorithm. The results show that, C-LRST performs better, with an average value of approximately $250ms$, while the values of CSGI, LRST-1, and LRST-2 are more than $500ms$.
% In terms of RTT jitter frequency and range, C-LRST has the lowest jitter frequency. The other three algorithms have stages of frequent oscillation, and the RTT jitter range exceeds $100ms$.
The reason for the observed RTT results of TCP is consistent with the aforementioned Ping.}

% haohao 7.7
% 我们选取了35组不同间距的地面站对并进行了end-to-end path analysis，从而让我们对四种算法有一个更全面的理解，并且make the findings more robust。
% 如图所示，散点表示不同距离的节点对在对应的算法下的Ping RTT和TCP RTT的平均值，直线为散点的线性拟合。整体的RTT会随着距离的增大而逐渐升高，一些反常是由于不同链路的切换次数不同。通过线性拟合可以看出，我们提出的C-LRST算法具有普遍最低的RTT，LRST-1和LRST-2较高，而CSGI算法最高。
\added{Then, we select 35 pairs of ground stations with different distances and performe end-to-end path analysis to provide a more comprehensive understanding of the four algorithms and make our findings more robust.}

\begin{figure}[!htp]
    \centering
    \subfigure[Ping RTT]{
        \includegraphics[width=0.46\linewidth]{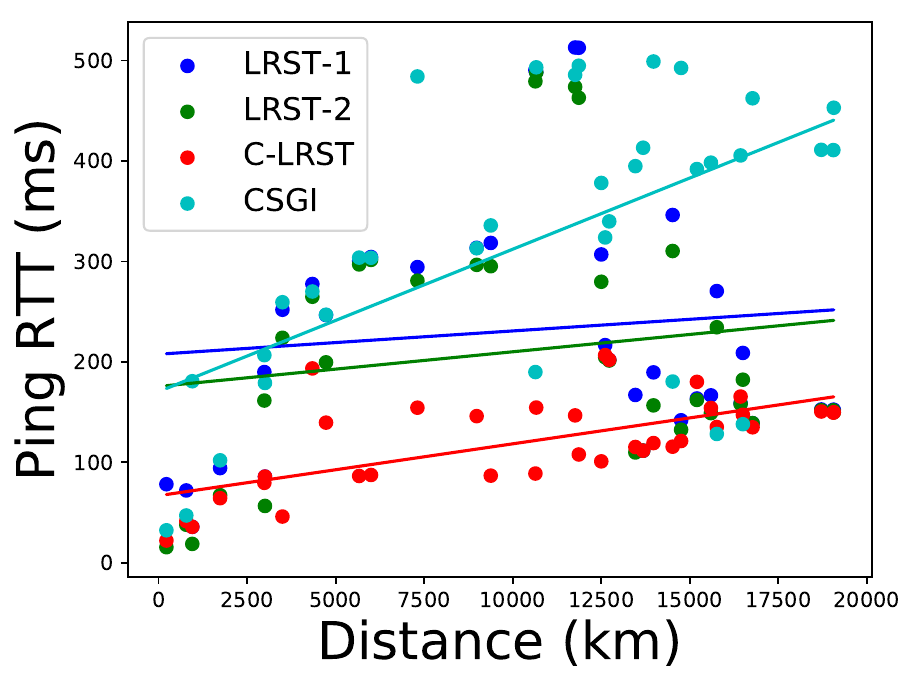}
        \label{subfig:ping_all}
    }
    \subfigure[TCP RTT]{
        \includegraphics[width=0.46\linewidth]{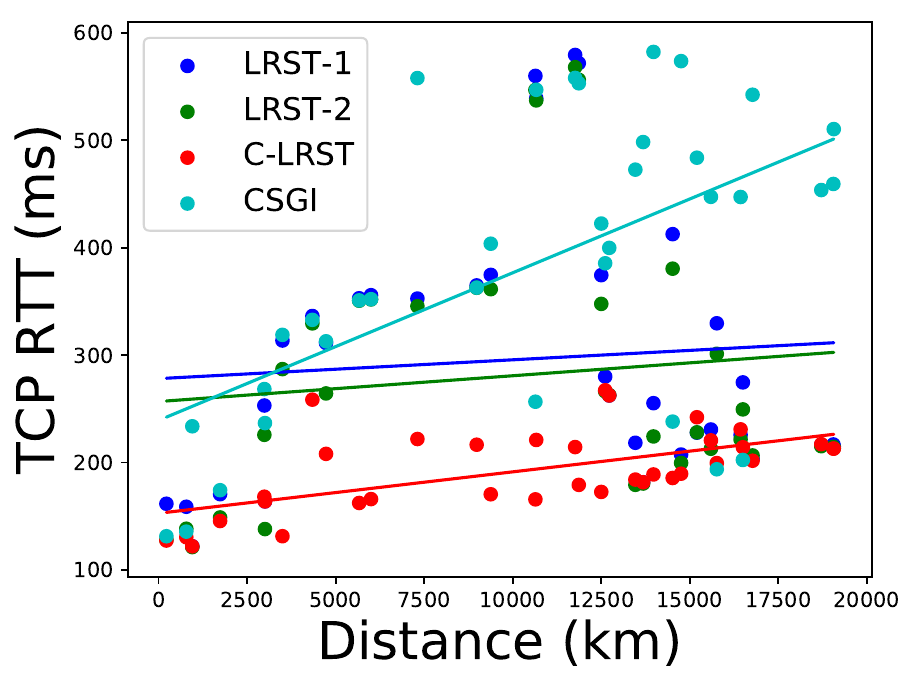}
        \label{subfig:tcp_all}
    }
    \caption{\added{Average RTT over distance between ground station pairs. Overall, C-LRST has the lowest average RTT, while CSGI has the highest.}}
    \label{fig:multi_paths}
\end{figure}

\added{As shown in Fig.~\ref{fig:multi_paths}, the scattered points represent the average RTT of pairs with different distances under the corresponding algorithm, and the straight line is the linear fit of the scattered points. The overall RTT gradually increases with the increasing distance, and some anomalies are due to the different switching numbers of different paths. From the linear fit, it can be seen that C-LRST  has the lowest RTT in general, LRST-1 and LRST-2 are higher, and CSGI is the highest.}

\added{\subsection{Scalability}}
% 为了研究提出的C-LRST算法的可扩展性，我们在更大的网络规模下进行了实验。
% 在原先星座规模的基础上，我们倍增了轨道面数，从而得到了原来两倍卫星数量的星座规模。
% 我们对比了原先星座规模和扩大的星座规模中不同距离的节点对的通信性能。如图x所示，图上的散点代表两个地面站节点在1000s的通信仿真过程中的平均时延，两条直线是散点的线性拟合。
% 两种星座规模的Ping RTT和TCP RTT的值基本相近，这是符合预期的。在星座构型不变的情况下，即使星座规模扩大，地面上两个节点通过卫星构建的通信链路也是接近的，因此会有相近的通信性能。然而，随着卫星数量的增多，更大规模的星座可能会带来额外的gsl切换。因此通过线性拟合可以看到，在远距离节点对通信场景下，扩大规模星座的两项RTT值有略微增大的趋势，这也是符合预期的。
% 通过对比实验可以看出，我们的C-LRST算法在更大规模的场景下仍然具有很好的适应性，尽可能地控制了星座规模扩大带来的更频繁的gsl切换问题，具有一定的可扩展性。
\added{In order to study the scalability of the proposed C-LRST algorithm, we conduct experiments under a larger-scale constellation. Based on the basic scale, we double the number of orbits, thereby obtaining a constellation with twice the number of satellites.}

\begin{figure}[!htp]
    \centering
    \subfigure[Ping RTT]{
        \includegraphics[width=0.46\linewidth]{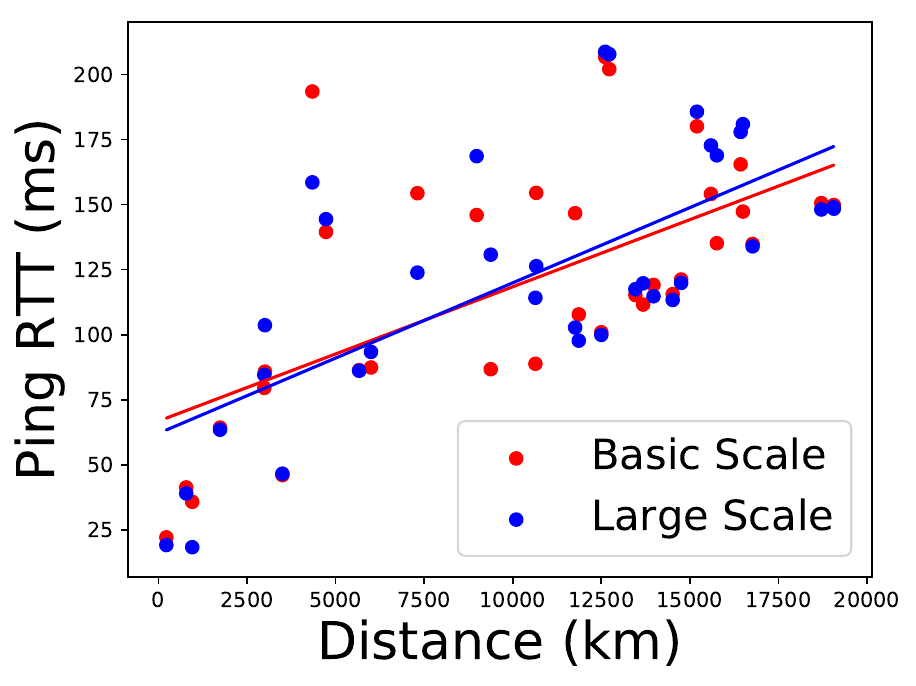}
        \label{subfig:new_starlink_ping_compare}
    }
    \subfigure[TCP RTT]{
        \includegraphics[width=0.46\linewidth]{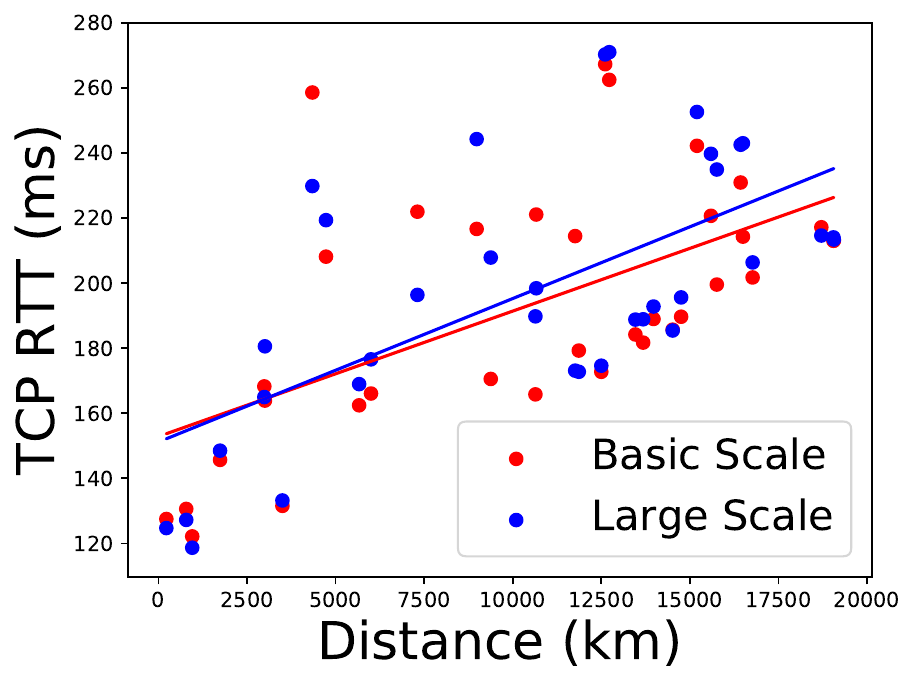}
        \label{subfig:new_starlink_tcp_compare}
    }
    \caption{\added{Average RTT of C-LRST over distance between ground station pairs under two scale constellations. Compared with the basic scale, C-LRST still maintains good performance under a larger-scale constellation.}}
    \label{fig:new_scale_compare}
\end{figure}

\added{We compare the communication performance of station pairs with different distances in the basic scale and the large scale constellations. As shown in Fig.~\ref{fig:new_scale_compare}, the scattered points represent the average RTT of ground station pairs during $1000s$ communication simulation, and the two straight lines are the linear fits of the scattered points.}

\added{The values of Ping RTT and TCP RTT of the two scale constellations are similar, which is expected. Even if the constellation scale is expanded, the communication links between the two nodes are similar, and thus there will be similar performance. However, with an increase in the number of satellites, a larger constellation may result in additional GSL switchings. Therefore, through linear fitting, it can be seen that in long-distance scenarios, the two values of the large scale constellation have a slightly increasing trend, which is also in line with expectations.}

\added{Through comparative experiments, it can be seen that our C-LRST algorithm still has good adaptability in larger-scale constellation, and controls the more GSL switchings problem caused by the expansion of the constellation scale as much as possible, and has certain scalability.}

% haohao 7.7

% haohao 7.7
\subsection{Overhead and deployment}
% \subsubsection{Overhead and deployment}
The computational complexity of the LRST algorithm depends on the operational complexity of calculating the longest service time satellites, so it is $O(n)$, where $n$ is the number of visible satellites from the ground terminal. Since the classification is based on the flight direction of the satellite, this information can be directly obtained from the TLE data, so the complexity is $O(n)$. Although the complexity of CSGI is low during handover, the computational complexity is exponential when initializing the satellite connected to the ground station at the initial moment. 
% \deleted{Therefore, it cannot be deployed normally in scenarios with hundreds of ground stations.}
% 在实际的部署中，CSGI算法需要掌握全部地面站的信息从而进行协调调度，在地面站数量很多的场景下难以部署。然而，我们的算法的选星策略是针对每个单独的地面站而言，它仅需要知道可以提供最长剩余服务时间的卫星，每个地面站的硬件条件能够承担这样的计算负载。另外，我们算法的设计基于对现有基础设施的调研，正如在Introduction提到的，现有的地面站如Starlink可以提供多接入能力，因此，我们的算法与现有的基础设施有较好的兼容性。
\added{In actual deployment, CSGI needs to grasp the information of all ground stations for coordinated scheduling, which is difficult to deploy in a scenario with a large number of ground stations. However, the satellite selection strategy of our algorithm is for each individual ground station, it only needs to know the satellite that can provide the longest remaining service time. The hardware conditions of each ground station can bear such a computing load. In addition, the design of our algorithm is based on an investigation of the existing infrastructure. As mentioned in the Introduction, the existing ground stations, such as Starlink, can provide multi-access capabilities. Therefore, our algorithm has good compatibility with existing infrastructure.}

% 需要指出的是，在更大数量的卫星场景下，地面站有更多的可见卫星，C-LRST的计算方法仍会产生一定的开销。然而，我们想要解释的是，这一计算开销的考虑是受限于我们的实验环境，计算可见卫星的服务时间本质上是因为地面站需要了解卫星的信息。在实际场景中，星座中的卫星位置，状态等信息会受到实时的监控和反馈，地面站会很容易获悉上空最长服务时间的卫星，并且由于卫星运动的规律性，也能容易预测下一时刻最长服务时间的卫星。因此，考虑到实际场景和实验环境的差异，实际部署的开销和效率应该是可接受的，即便是在更大规模的卫星和地面站场景下。
\added{It should be pointed out that in the scenario of a larger number of satellites, there are more visible satellites on the ground station, and the calculation method of C-LRST will still generate certain costs. However, we would like to explain that this consideration of cost is limited to our experimental environment, and the calculation of visible satellite service time is essentially because the ground station needs to know the satellites information. However, in actual scenarios, the position, status, and other information of the satellites in the constellation will be monitored and fed back in real time. The ground station will easily know the satellite with the longest service time, and it will be easy to predict the satellite with the longest service time in the next moment owing to the regularity of the satellite movement. Therefore, considering the differences between the actual scenario and the experimental environment, the overhead and efficiency of the actual deployment could be acceptable, even in a scenario with a large number of ground stations or satellites.}

% 与真实世界的网络系统不同，受限于实验条件和仿真能力的限制，我们的实验和性能评估存在天然的瓶颈。但是，我们的研究主要是为了贡献一种星地互联的设计方案，为未来的星座建设提供一定的参考，在现实场景中的应用需要考虑各种实际因素进行相应的调整和完善。
\added{Unlike real-world network systems, there are natural bottlenecks in our experiments and performance evaluations due to the limited experimental conditions and simulation capabilities. However, our research is mainly aimed at contributing a design scheme for satellite-ground interconnection and providing a reference for future constellation construction. The application in a real scenario needs to consider various practical factors for corresponding adjustment and improvement.}

\section{Conclusion}
\label{sec:conclusion}
In this paper, we propose the Classification-based Longest Remaining Service Time (C-LRST) algorithm. C-LRST supports the actual scenario with multi-access capabilities for terminals, and based on this, it adds more optional paths during routing with low computational complexity. 
In addition, we conduct experiments to evaluate our algorithm on the open-source constellation simulation platform Hypatia. We make adaptive improvements to the Hypatia platform to support our algorithm. Experiment results show that compared with current algorithms, the network delay of C-LRST is reduced by about 60\%, the average throughput is increased by about 40\%, and the link is more stable. Considering the access capabilities of current ground terminals, our algorithm is more suitable for application scenarios. As a potential future direction, we are looking forward to extending our C-LRST to improve the performance of various applications such as
distributed learning systems~\cite{lin2024split,hu2023holofed,zheng2023autofed,hu2024accelerating,lin2024adaptsfl}, large language models~\cite{lin2024splitlora,fang2024automated,qiu2024ifvit}, mobile edge computing~\cite{gu2024loccams,lin2022tracking,zheng2021more,lin2022v2i,chen2021rf,fang2024ic3m,wu2024s,wu2024rethinking}
etc in LEO satellite networks.

\bibliography{reference}
\bibliographystyle{IEEEtran}

\end{document}